\documentclass[11pt,a4paper]{article}
\pdfoutput=1
\usepackage{jheppub}

\newcommand{\half}{{\textstyle\frac{1}{2}}}
\newcommand{\be}{\begin{eqnarray}}
\newcommand{\ee}{\end{eqnarray}}
\newcommand{\nn}{\nonumber\\}

\newcommand{\lsim}{\;\raisebox{-0.9ex}{$\textstyle\stackrel{\textstyle<}{\sim}$}\;}

\def\delr            {\!\stackrel{\leftrightarrow}{\partial^\mu}\!}
\def\a               {\alpha}
\def\b               {\beta}
\def\l               {\lambda}

\title{CP violation in charged Higgs production and decays\\ in the Complex Two~Higgs Doublet Model}

\author[a,b]{A. Arhrib}
\author[c]{E. Christova}
\author[d]{H. Eberl}
\author[d,e]{E. Ginina}

\affiliation[a]{D\'epartement de Math\'ematique, Facult\'e des
Sciences et Techniques, Universit\'e Abdelmalek Essa\^adi,\\
B.~416, Tangier, Morocco}
\affiliation[b]{LPHEA, Facult\'e des Sciences-Semlalia, \\
B.P.~2390 Marrakesh, Morocco}
\affiliation[c]{Institute for Nuclear Research and Nuclear Energy, BAS,\\
Sofia 1784, Bulgaria}
\affiliation[d]{Institut f\"ur Hochenergiephysik der \"Osterreichischen Akademie der
Wissenschaften,\\
A-1050 Vienna, Austria}
\affiliation[e]{Universit\"at Wien, Fakult\"at f\"ur Physik, \\ A-1090 Vienna, Austria}

\emailAdd{aarhrib@ictp.it}
\emailAdd{echristo@inrne.bas.bg}
\emailAdd{helmut@hephy.oeaw.ac.at}
\emailAdd{eginina@hephy.oeaw.ac.at}

\abstract{We study the effects of CP violation in charged Higgs boson production $pp\to tH^\pm + X$
at the LHC, as well as in the charged Higgs boson decays $H^\pm \to tb$ and $H^\pm \to
W^\pm H_i^0$, $i=1, 2, 3$. The study is done in the framework of the type II complex Two Higgs Doublet
Model (2HDM) with softly broken $Z_2$ symmetry. In this model violation of CP invariance is induced by the
complex parameter $m^2_{12}$ of the tree-level Higgs potential. We calculate the
CP violating rate asymmetries for $H^+$ and $H^-$
production and decays as well as for the combined processes at one-loop level
and perform a detailed numerical analysis. All calculations are done with the automatic amplitude generator FeynArts
and the calculational tool FormCalc, for which we have written a complete complex 2HDM model file and
relevant fortran drivers. The implementation of the
complex 2HDM
in FeynArts and FormCalc is described. In comparison with the
analogous results in the MSSM, all considered CP violating asymmetries are smaller
by an order of
magnitude and do not exceed $2 \div 3\%$.}


\arxivnumber{1011.6560}

\begin{document}

\maketitle
\flushbottom

\section{Introduction}
\label{sec:intro}

The CERN Large Hadron Collider (LHC) has started its operation aiming at a direct
verification of one of the different
candidates that generalize the Standard Model (SM). The
experiments at the LHC have to distinguish between the predictions of the various theoretical models. Having in
mind the large number of free parameters that most of these models
introduce, this task is highly nontrivial. Powerful software and
hardware resources are required so that their analyses lead to
definite predictions for experimental searches.

An important approach for testing fundamental theories is studying
discrete symmetries as model properties.
In particular, the CP symmetry is known to be violated in
nature \cite{CP&bariogenesis,CP&bariogenesis2,CP&bariogenesis3}. According to Sakharov's
theorem, the mechanism of CP violation (CPV) in the SM is not strong enough
to explain the baryon asymmetry in the universe -- more
CPV is needed. For that reason all models beyond the SM suggest
additional and different sources of CPV.

On the other hand, the Higgs boson is not yet found and the mechanism of electroweak
symmetry breaking (EWSB)
remains the only part of the SM which is not yet verified.
The extensions of the SM enlarge the Higgs sector and predict both
charged and new neutral Higgs bosons.
If a neutral Higgs boson is
discovered at the LHC, there will be a long way to determine wether it belongs to the SM or
to some of its extensions. A discovery of a charged Higgs boson though, would be a clear
signal for Physics beyond the SM. A possibility to distinguish between the different
models containing a charged Higgs is looking for effects of
CPV in processes with $H^\pm$, which we shall explore in this paper.

The simplest extensions of the SM are the models with two Higgs doublets.
All of them provide new sources of CPV.
The most popular one is the Minimal Supersymmetric Standard Model
(MSSM). The MSSM Higgs sector is a constrained 2HDM of type II.
On the other hand, the general Complex Two Higgs Doublet Model (C2HDM) has attracted much
attention due to the CPV it can
accommodate \cite{THDM, THDM2, THDM3} and due to its simplicity.
The physical mass eigenstates in the Higgs sector of the 2HDM and the MSSM are the same
-- there are two charged $H^\pm$  and three
neutral $H^0_i$, $i=1,2,3$ Higgs bosons. Processes involving $H^\pm$ can generate a CPV asymmetry at one-loop
level in both MSSM and C2HDM. However, the CPV sources in these two models are different.

In the MSSM, the tree-level Higgs potential is real and thus the
neutral Higgs bosons have definite CP parities and preserve
CP invariance. CPV results from the non-zero
CP phases of the higgsino mass parameter $\mu=|\mu|~e^{i\phi_\mu}$ in the
superpotential, the gaugino mass parameters $M_i=|M_i|~e^{i
\phi_{i}},\, i=1,2,3$, and the trilinear couplings $A_f=|A_f|~e^{i
\phi_f}$ ($f$ stands for a fermion) in the soft SUSY breaking
Lagrangian. In particular, in the MSSM neutral Higgs sector,
 the presence of these CP violating phases induces mixing
between the CP-even and the CP-odd scalars at one-loop level, yielding three
mass eigenstates \cite{Carena}.

In the C2HDM  CPV is induced by
the complex parameters of the tree-level Higgs potential and
thus the physical neutral Higgs bosons
are mixtures of the CP-even and the CP-odd states.
The interactions of these neutral Higgs
bosons with fermions and gauge bosons violate CP invariance.
We stress that in the C2HDM  the mixture of the
CP-even and the CP-odd neutral Higgs bosons is already at tree-level, while
in the MSSM it
is a loop-induced effect, generated by SUSY-loop corrections.

Effects of CPV in the decays of $H^\pm$ into "ordinary" (SM) particles in the
framework of the MSSM were studied in \cite{Christova:2002ke,Christova:2002ke2,Christova:2003hg,Hollik:2010dh}.
These CPV decay rate asymmetries are of
interest for a future linear collider, where the charged Higgs
will be produced in pairs. There CPV occurs only in the decays of
$H^\pm$. For the LHC one must take into account CPV in
the production process as well.
Recently, $W^\pm H^\mp$ production at the LHC was studied in \cite{Dao:2010nu}.
In \cite{CPVinH+tMSSM,CPVinH+tMSSM2,CPVinH+tMSSM3} we studied CPV in the combined process of
a charged Higgs boson production at the LHC:
\be
pp~\to ~H^\pm ~t ~+~X\,,\label{prod}
\ee
followed by  the subsequent decays of $H^\pm$, where
the production process (\ref{prod}) is due to the partonic
process $bg\to H^\pm\, t$. In both production and decays
CPV is induced by one-loop radiative corrections with
supersymmetric (SUSY) particles
in the loops. Our numerical analysis showed that the
CPV asymmetries, both in the decays and in the production, can be rather
large. However, in the $H^\pm \to t\bar{b}$ decay mode, there can
occur cancellations reducing the total asymmetry.

In this paper we perform an analogous study within the complex 2HDM. In addition to the production
rate asymmetry we calculate the CPV decay rate asymmetries
in the dominant decay modes of $H^\pm$ in the C2HDM:
\be
H^\pm \to t  b\quad  {\rm and}\quad H^\pm \to W^\pm H^0_i,\quad i=1,2\,.\label{decays}
\ee
We also consider the CPV asymmetries
in the combined processes of production  (\ref{prod}) and decays  (\ref{decays}).
We work in type II C2HDM with softly broken $Z_2$ symmetry of the Lagrangian.
In order to perform a numerical analysis, we have
generalized the existing codes of the FeynArts (FA) and
FormCalc (FC) packages \cite{FeynArts,FeynArts2,FeynArts3,FeynArts4}, for calculating processes in the C2HDM.

The paper is organized as follows. In
section \ref{sec:2HDM} we shortly review the
general 2HDM, fix the conventions for the scalar potential, the Yukawa interactions
and the parameter set we work with.
We derive the Higgs couplings
to fermions and gauge bosons and list the existing
theoretical constraints on the 2HDM Lagrangian.
In the third section we study the production and decay processes mentioned above,
present the expressions for the CPV asymmetries, and show the one-loop
contributions to these asymmetries in the C2HDM.
We proceed with a detailed numerical analysis in section \ref{sec:numerics}. The numerical results for
the CPV asymmetries in the charged Higgs decays  (\ref{decays}), in the charged Higgs production  (\ref{prod}), as well
as for the combined process of production  (\ref{prod}) and subsequent decays  (\ref{decays})
at LHC at the center-of-mass energy $\sqrt{s}=14~$TeV in the C2HDM are presented.
We also discuss the experimental constraints on the C2HDM parameter space.
The numerical analysis is done using the FA and FC packages. The implementation of the C2HDM into these packages is given
in detail in the appendix, which also contains
the Lagrangian relevant for our study, the expression for $\Delta \rho$ implemented in the code and some important
relations between the parameters of the scalar potential and
the physical Higgs masses. As usual, the paper ends with conclusions.

\section{General 2HDM: short review and notations}
\label{sec:2HDM}

\subsection{Scalar potential and its parameterization}
\label{sec:genpot}

The general 2HDM is obtained via extending the SM Higgs sector,
consisting of one complex Y= +1,
$\rm{SU (2)_L}$ doublet scalar field $\Phi_1$, with an additional complex
Y = +1, $\rm{SU (2)_L}$ doublet scalar field $\Phi_2$.
Using $\Phi_{1,2}$, one can build the most
general renormalizable $\rm{SU(2)_L \times U(1)_Y}$ gauge
invariant Higgs potential \cite{THDM_CPC1,THDM_CPC2,THDM_CPC22,THDM_CPC23,tdlee,tdlee2,tdlee3,tdlee4}:
\begin{eqnarray}
\hspace*{-1cm}
 V_{\rm{Higgs}}(\Phi_1,\Phi_2) = \frac{\l_1}{2}(\Phi_1^\dagger\Phi_1)^2 +
\frac{\l_2}{2}(\Phi_2^\dagger\Phi_2)^2 +
\l_3(\Phi_1^\dagger\Phi_1)(\Phi_2^\dagger\Phi_2) +
\l_4(\Phi_1^\dagger\Phi_2)(\Phi_2^\dagger\Phi_1) \nn  +~
\frac12\left[\l_5(\Phi_1^\dagger\Phi_2)^2 +\rm{h.c.}\right]
+~\left\{\left[\l_6(\Phi_1^\dagger\Phi_1)+\l_7(\Phi_2^\dagger\Phi_)\right]
(\Phi_1^\dagger\Phi_2)+\rm{h.c.}\right\} \nn
-~\frac{1}{2}\left\{m_{11}^2 \Phi_1^\dagger \Phi_1+ \left[m_{12}^2
\Phi_1^\dagger \Phi_2 + \rm{h.c.}\right]
 + m_{22}^2\Phi_2^\dagger
\Phi_2 \right\}\,.\label{CTHDMpot}
\end{eqnarray}
By hermiticity of eq. (\ref{CTHDMpot}),
$\l_{1,2,3,4}$, as well as $m_{11}$ and $m_{22}$ are real-valued;
while the dimensionless parameters $\l_{5}$, $\l_6$, $\l_7$ and $m_{12}^2$ are in general complex.

\subsubsection{Mass eigenstates}

After the $\rm{SU(2)_L \times U(1)_Y}$ gauge symmetry is broken down
to $\rm {U (1)_{em}}$ via the Higgs mechanism, one can choose a basis
where the vacuum expectation values (VEVs) of the two Higgs doublets, $v_1$ and $v_2$ are
non-zero, real and positive, and fix the following
parameterization \cite{Maria,Per}:
\be
 \Phi_1= {\small \left(
  \begin{array}{c}
   \varphi_1^+\\ (v_1+\eta_1+i\chi_1)/\sqrt{2}
  \end{array}\right)}\,,
\quad \Phi_2= {\small \left( \begin{array}{c}
  \varphi_2^+\\ (v_2+\eta_2+i\chi_2)/\sqrt{2}
  \end{array}\right)}\,.\label{Hdoublets}
\ee
Here $\eta_{1,2}$ and $\chi_{1,2}$ are neutral scalar fields
and $\varphi_{1,2}^\pm $ are charged scalar fields.
The physical Higgs eigenstates are obtained as follows.

The charged Higgs fields $H^\pm$ and the
charged would-be Goldstone boson fields $G^\pm$ are a mixture of the
charged components of the Higgs doublets  (\ref{Hdoublets}), $\varphi^\pm_{1,2}$:
\begin{eqnarray}
H^\pm &=& -\sin\beta\varphi^\pm_1+\cos\beta\varphi^\pm_2\,,
\nn
G^\pm &=& \cos \beta \varphi^\pm_1+\sin\beta\varphi^\pm_2\,,
\end{eqnarray}
where the mixing angle $\beta$ is defined through the ratio of the
VEVs of the two Higgs doublets $\Phi_2$ and $\Phi_1$, $\tan\beta=v_2/v_1$. $G^\pm$ give masses to the
$W^\pm$ bosons.

Obtaining the neutral physical Higgs states is a few steps procedure.
First, one rotates the
imaginary parts of the neutral components of eq. (\ref{Hdoublets}): $(\chi_1,\chi_2)$
into the basis $(G^0,\eta_3)$:\footnote{Note that in the case of
$m_{12}^2=\l_{6}=\l_7=0$ and all other parameters of eq. (\ref{CTHDMpot}) are real,
the physical Higgs sector of the 2HDM is analogous to the one of
the tree-level MSSM. In this case the scalar field $\eta_3$
is equivalent to the  MSSM neutral CP-odd Higgs boson $A^0$.}
\begin{eqnarray}
G^0 &=& \cos \beta \chi_1+\sin\beta\chi_2\,, \nn
\eta_3 &=& -\sin\beta\chi_1+\cos\beta\chi_2\,,
\label{CPoddHiggs}
\end{eqnarray}
where $G^0$ is the would-be Goldstone boson which
gives a mass to the $Z$ gauge boson.
After elimination of the Goldstone mode, the remaining
neutral CP-odd component $\eta_3$ mixes with the neutral CP-even
components $\eta_{1,2}$. The relevant squared mass
matrix ${\cal M}^2_{ij}=\partial^2 V_{\rm{Higgs}}/(\partial
\eta_i\partial \eta_j )$, $i,j=1,2,3$,
has to be rotated from the so called "weak basis" $(\eta_{1},\eta_{2},\eta_{3})$ to the diagonal basis
$(H_1^0, H_2^0, H_3^0)$ by an orthogonal $3\times 3$ matrix ${\cal{R}}$ as follows:
\be
{\cal R} {\cal M}^2 {\cal R}^T={\cal
M}^2_{\rm{diag}}={\rm diag}({\rm M}_{H_1^0}^2, {\rm M}_{H_2^0}^2, {\rm M}_{H_3^0}^2)\,, \label{mixmat}
\ee
with
\be
{\small \left(
  \begin{array}{c}
   H_1^0\\ H_2^0 \\H_3^0
  \end{array}\right)}=
{\cal R} {\small \left( \begin{array}{c}
  \eta_1\\ \eta_2 \\ \eta_3
  \end{array}\right)}\,,
\ee
where we have defined the Higgs fields $H_i^0$ such that their masses satisfy the
inequalities:
\be
\rm{M_{H_1^0}}\leq\rm{M_{H_2^0}}\leq\rm{M_{H_3^0}}\,.
\ee
Note that the mass eigenstates $H_i^0$ have a mixed CP structure.

Following \cite{ElKaffas},
we parametrize the orthogonal $3\times 3$ matrix
${\cal{R}}$ by three rotation angles ~$\alpha_i,~
i=1,2,3$:
\begin{eqnarray}
\cal{R} &=& {\small \left(
  \begin{array}{ccc}
   1         &    0         &    0
\\ 0 &  \cos\alpha_3 & \sin\alpha_3 \\ 0 & -\sin\alpha_3 &
\cos\alpha_3
  \end{array}\right)}
 {\small \left(
  \begin{array}{ccc}
\cos\alpha_2 & 0 & \sin\alpha_2 \\ 0         &       1         & 0
\\ -\sin\alpha_2 & 0 & \cos\alpha_2
  \end{array}\right)}
   {\small \left(
  \begin{array}{ccc}
\cos\alpha_1 & \sin\alpha_1 & 0 \\ -\sin\alpha_1 & \cos\alpha_1 &
0 \\ 0         &       0         & 1
  \end{array}\right)}\nn
 &=& {\small \left(
  \begin{array}{ccc}
   c_1\,c_2 & s_1\,c_2 & s_2 \\ - (c_1\,s_2\,s_3 + s_1\,c_3) & c_1\,c_3 -
s_1\,s_2\,s_3 & c_2\,s_3 \\ - c_1\,s_2\,c_3 + s_1\,s_3 & -
(c_1\,s_3 + s_1\,s_2\,c_3) & c_2\,c_3
  \end{array}\right)}\,,
  \label{matRo}
\end{eqnarray}
with $s_i=\sin\alpha_i$ and $c_i=\cos\alpha_i$, which we vary in our numerical analysis in the following ranges:
\be
-\frac{\pi}{2} < \a_1\leq \frac{\pi}{2}~; \quad -\frac{\pi}{2} <
\a_2\leq \frac{\pi}{2}~; \quad 0 \leq \a_3\leq \frac{\pi}{2}~.
\ee

\subsection{$Z_2$ symmetry and input parameter set}
\label{sec:Z2param}

In the most general 2HDM, some types of Yukawa interactions
can introduce flavour changing neutral currents
(FCNC) already at tree level. It is well known that the latter effects are small in nature.
This problem has been solved by imposing a discrete $Z_2$ symmetry on the Lagrangian. It forbids
 $\Phi_1\leftrightarrow\Phi_2$ transitions and in its exact form it also leads
 to conservation of CP \cite{Glashow}. In order to allow some effects of CPV
it is necessary to violate the $Z_2$ symmetry. Basically, there are two ways of $Z_2$ symmetry violation -- "soft"
 and "hard". A softly broken $Z_2$ symmetry suppresses FCNC at tree level, but still allows CPV.

In this paper we will work in a model of a softly broken $Z_2$ symmetry of the 2HDM Lagrangian.
This forbids the quartic terms proportional to $\l_6$ and $\l_7$ in eq. (\ref{CTHDMpot}), but the quadratic term with $m^2_{12}$ is still allowed
 \cite{Khater, Per2}:
\begin{eqnarray}
\hspace*{-2cm}
 V_{\rm{Higgs}}^{\rm soft}(\Phi_1,\Phi_2) = \frac{\l_1}{2}(\Phi_1^\dagger\Phi_1)^2 +
\frac{\l_2}{2}(\Phi_2^\dagger\Phi_2)^2 +
\l_3(\Phi_1^\dagger\Phi_1)(\Phi_2^\dagger\Phi_2) +
\l_4(\Phi_1^\dagger\Phi_2)(\Phi_2^\dagger\Phi_1) \nonumber \\  +~
\frac12\left[\l_5(\Phi_1^\dagger\Phi_2)^2 +\rm{h.c.}\right]
-~\frac{1}{2}\left\{m_{11}^2 \Phi_1^\dagger \Phi_1+ \left[m_{12}^2
\Phi_1^\dagger \Phi_2 + \rm{h.c.}\right]
 + m_{22}^2\Phi_2^\dagger
\Phi_2 \right\}\,.\label{THDMpot}
\end{eqnarray}

The Higgs potential ~(\ref{THDMpot}) has 12 real parameters: 2 real masses: $m^2_{11,\,22}$, 2 VEVs: $v_{1,\,2}$, four real quartic couplings: $\l_{1,\,2,\,3,\,4}$ and two complex parameters: $\l_5$ and $m_{12}^2$. The conditions for having an extremum of eq. (\ref{THDMpot}) reduce the number of parameters:
$m^2_{11\,,22}$ are eliminated by the minimization conditions, and the combination $v_1^2+v_2^2$ is fixed at the electroweak scale $v=(\sqrt 2 G_F)^{-1/2}$ = 246 GeV. Moreover, in this case the minimization conditions also relates ${\rm Im}\, (m_{12}^2)$ and ${\rm Im}\, (\l_5)$:
 \be
{\rm Im}\, (m_{12}^2) = v_1\,v_2\, {\rm Im}\,(\l_5)\,. \label{oneCP}
 \ee
Thus, our Higgs potential  (\ref{THDMpot}) is a function of 8 real independent parameters:
 \be
\big\{\l_{1,2,3,4},~~ {\rm Re}\, (\l_5),~~ {\rm Re}\,
(m_{12}^2),~~\tan \beta,~~ {\rm Im}\, (m_{12}^2)\big\}. \label{set1}
\ee
It contains minimal CPV generated by $m_{12}^2 \ne 0$ and complex.
In our further analysis we will use the following parameter set equivalent to eq. (\ref{set1}):
\be
\bigg\{M_{H_1^0},~~ M_{H_2^0},~~ M_{H^+},~~ \alpha_1,~~
\alpha_2,~~ \alpha_3,~~\tan \beta,~~ {\rm Re}\, (m_{12})
\bigg\}\,.\label{paramm}
\ee
Note that the mass of the heaviest neutral Higgs boson $H_3^0$ is not an independent parameter.
In the considered CP violating case, the matrix elements $({\cal M}^2)_{13}$
and $({\cal M}^2)_{23}$ of the squared mass
matrix  (\ref{mixmat}) are non-zero and correlated \cite{theor.constr.}:
\be
({\cal M}^2)_{13}&=&\tan \b \,({\cal M}^2)_{23}\,.
\label{sumrule0}
\ee
Writing this relation in terms of the physical masses
${\rm M}_{H_1^0}$, ${\rm M}_{H_2^0}$, ${\rm M}_{H_3^0}$ one obtains \cite{theor.constr.}:
\be
M_{H_3^0}^2=\frac{M_{H_1^0}^2 R_{13}(R_{12}\tan
\beta-R_{11})+M_{H_2^0}^2R_{23}(R_{22}\tan
\beta-R_{21})}{R_{33}(R_{31}-R_{32}\tan \beta)}\,,\label{M3}
\ee
where $ R_{ij}\,, i,j=1,2,3,$ are the elements of the rotation matrix (\ref{matRo}).

The expressions for the parameters $\l_{1,2,3,4},~ {\rm Re} \l_5, {\rm Im} \l_5$ of the scalar potential  (\ref{THDMpot}) as functions of the physical
masses and mixing angles are given in appendix \ref{App:lambda}.

In appendix \ref{App:Lag} we also list the triple scalar couplings of the Higgs bosons calculated
from the potential  (\ref{THDMpot}) and relevant to our study. Note, that the
couplings ${\cal C}({\rm H}_i^0~{\rm H}^\pm~ {\rm G}^\mp),\, i=1,2,3$, have both real and
imaginary parts and can lead to CPV.

\subsection{Higgs and gauge boson interactions}
\label{sec: higgsgauge}

The Higgs and gauge boson interactions arise from
the covariant derivatives of the doublet fields $\Phi_{1,2}$:
\be
{\cal L}_{\rm gauge}=\sum_{i=1}^{2} (D^\mu \Phi_i)^\dagger (D_\mu \Phi_i),
\quad  D^\mu =\partial^\mu +ig \vec{T}_a\vec{W}_\mu^a+ig'
Y_{i}B_\mu/2\label{covder}
\ee
where $\vec{T}_a$ are the isospin  generators, $Y_{i}$ are
the hypercharges of the Higgs $\Phi_i$, $\vec{W}_\mu^a$ and  $B_\mu$
are the ${\rm SU(2)_L}$ and  ${\rm U(1)_Y}$ gauge fields, $g$, $g'$
are the corresponding ${\rm SU(2)_L}$ and  ${\rm U(1)_Y}$ gauge couplings.

After having rotated the Higgs and gauge bosons fields to their
mass eigenstates bases one obtains terms of triple and quartic interactions
between them. The interactions relevant to our study are
${\rm H}_i^0{\rm W}{\rm W}$ and ${\rm H}_i^0~{\rm W}^\pm~{\rm H}^\mp$, with the corresponding couplings
\begin{eqnarray}
{\cal C}({\rm H}_i^0{\rm W}{\rm W}) &=& \cos\beta R_{i1} +\sin\beta R_{i2}\, ,\nn
{\cal C}({\rm H}_i^0~{\rm W}^\pm~{\rm H}^\mp) &=& \mp~ i( \sin\beta R_{i1}-\cos\beta R_{i2})\pm R_{i3}\, .
\label{gaugecoup}
\end{eqnarray}
Note that ${\cal C}({\rm H}_i^0{\rm W}{\rm W})$ is purely real, while
${\cal C}({\rm H}_i^0~{\rm W}^\pm~{\rm H}^\mp)$ has
both real and imaginary parts and
can lead to effects of CPV, as we will see later. The relevant Lagrangian is given in appendix \ref{App:Lag}.

From the condition for unitarity of the rotation matrix ${\cal R}$ one derives the following sum rules:
\begin{eqnarray}
{\cal C}({\rm H}_i^0{\rm W}{\rm W})^2 + |{\cal C}({\rm H}_i^0~{\rm W}^+~{\rm H}^-)|^2 &=& 1 \quad \rm{for\ each}\ i=1,2,3
\label{sum1}\\
\sum_{i=1}^3{\cal C}({\rm H}_i^0{\rm W}{\rm W})^2 &=& 1\label{sum2}
\end{eqnarray}
From eqs. (\ref{sum1}) and (\ref{gaugecoup}) follows that if for a fixed $i$
 the term $|{\cal C}({\rm H}_i^0~{\rm W}^+~{\rm H}^-)|^2$ is suppressed, then
 $(\sin\beta R_{i1}-\cos\beta R_{i2})^2\approx 0$ and
$R_{i3}^2\approx 0$. The relation  $R_{i3}^2\approx 0$ implies that
 $H_i^0$ has a very small pseudoscalar component and is
dominantly a CP-even  state. Furthermore, the
  sum rule  (\ref{sum2}) implies that the other Higgs states decouple, i.e.
   ${\cal C}({\rm H}_j^0{\rm W}{\rm W})^2\approx 0$ for $j\neq i$.
We will come back to the physical consequences of these sum rules
in our physics discussion on the studied decay and production processes in section \ref{sec:numerics}.

\subsection{Yukawa interactions}
\label{sec:yuk}

In the framework of the 2HDM various models of Yukawa interactions can be realized \cite{THDM}. Depending on the Yukawa interaction
one distinguishes between different types of 2HDM's.

The most general 2HDM, in which each fermion doublet and singlet  couples to both Higgs doublets is called the
type-III model. This model leads to FCNC already at tree level and is a
subject of severe constraints from flavour physics
observables \cite{THDM_CPC2,WahabElKaffas:2007xd}.
As we already discussed, in order to avoid problems with FCNC usually a
$Z_2$ discrete symmetry is imposed on the Lagrangian \cite{Glashow}.

In the 2HDM type-I all fermions, quarks and leptons, couple to only one of
the Higgs doublets exactly like in the SM. These models are interesting as the
decoupling Higgs is a natural candidate for dark matter.

There exists another type of 2HDM where one of the Higgs doublets
couples to leptons and the other one to the up and down
quarks \cite{YukTHDM1,YukTHDM12,YukTHDM13}. The phenomenology of these models has been
reviewed recently in \cite{YukTHDM2,YukTHDM22,YukTHDM23}.

In the present paper we work in type-II 2HDM. In this model,
down-type quarks and charged leptons couple to
$\Phi_1$ and the up-type quarks couple to the other Higgs doublet
$\Phi_2$.
( In the MSSM the Higgs sector is also a 2HDM type-II, but with the Higgs self-interactions fixed
by gauge couplings. )
In order to avoid FCNC at tree level, but allow for CPV,
our discrete $Z_2$ symmetry is softly broken, i.e.
$\l_6=\l_7=0$, but $m_{12}^2$ is non-zero and complex,
see section \ref{sec:Z2param}.

The parts of the interaction Lagrangian of Higgs bosons and fermions relevant for our discussions are given in
appendix \ref{App:Lag}.
Since the neutral Higgses are mixtures of
CP-odd and CP-even states, their couplings
to a fermion pair have the general form
 $a+i~b~\gamma_5$ with $a$ and $b$ real, that can lead to CPV.

\subsection{Theoretical constraints on the Higgs potential}
\label{sec:th.constr.}

The Higgs potential $V_{\rm Higgs}^{\rm soft}$ given
by eq. (\ref{THDMpot}) has to satisfy some general requirements
like positivity, unitarity and perturbativity \cite{ElKaffas}. These requirements
together with the minimum conditions naturally lead to constraints on its parameters \cite{Maria, ElKaffas, theor.constr.}. The theoretical
constraints on the potential  (\ref{THDMpot}) are well studied
and will not be a subject of this note. For completeness, we list below the expressions
we have implemented in our numerical code.

In order to have a stable vacuum the potential should be
positive for large values of $|\phi_k|$, which leads to the constraints \cite{Maria}:
\be
\lambda_1>0\,, \quad \lambda_2>0\,, \quad \lambda_3+
\sqrt{\lambda_1 \lambda_2}>0\,, \quad \lambda_4+\lambda_4-|\lambda_5|+
\sqrt{\lambda_1 \lambda_2}>0\,.\label{th.constr.1}
\ee
From the requirement that theory remains perturbative we have:
$|\lambda_i| \leq 8 \pi$ for any $i$. The unitarity requirement means
that the tree-level amplitudes for Higgs-Higgs, Higgs-vector boson
and vector boson-vector boson scattering should not exceed the
unitarity limit when contributing to the s-wave.
Both unitarity and perturbativity requirements
lead to the constraints \cite{Maria}:
\begin{eqnarray}
\left|~\frac{1}{2}\left( \lambda_1+\lambda_2\pm \sqrt{(\lambda_1-\lambda_2)^2+4 |\lambda_5|^2}\right)\right| &<& 8\pi\,, \nn
\left|~\frac{1}{2}\left(\lambda_1+\lambda_2\pm \sqrt{(\lambda_1-\lambda_2)^2+4\lambda_4^2} \right) \right| &<& 8 \pi\,, \nn
\left|~\frac{3(\lambda_1+\lambda_2)\pm\sqrt{9(\lambda_1-\lambda_2^2+4(\lambda_3+\lambda_4)^2)} }{2} \right| &<& 8 \pi\,, \nn
|~\lambda_3\pm\lambda_4|<8 \pi\,, \quad |~\lambda_3\pm |\lambda_5||<8 \pi\,,\quad |~\lambda_3+2 \lambda_4\pm|\lambda_5|| &<& 8 \pi\,.
\label{th.constr.2}
\end{eqnarray}
As the eqs.~(\ref{th.constr.1}) and (\ref{th.constr.2}) only depend on the absolute value of
$\l_5$, they do not constrain the phase of $\l_5$.
We will see later, that these theoretical constraints already strongly reduce the
C2HDM parameter space \cite{Maria, ElKaffas, theor.constr.}.

\section{CP violation in $H^\pm$ production and decays}
\label{sec:applic}

\subsection{The processes}\label{sec:pro}

\begin{figure}[h!]
\begin{center}
\vspace*{-1.8cm}
{\mbox{\resizebox{!}{18.0cm}{\includegraphics{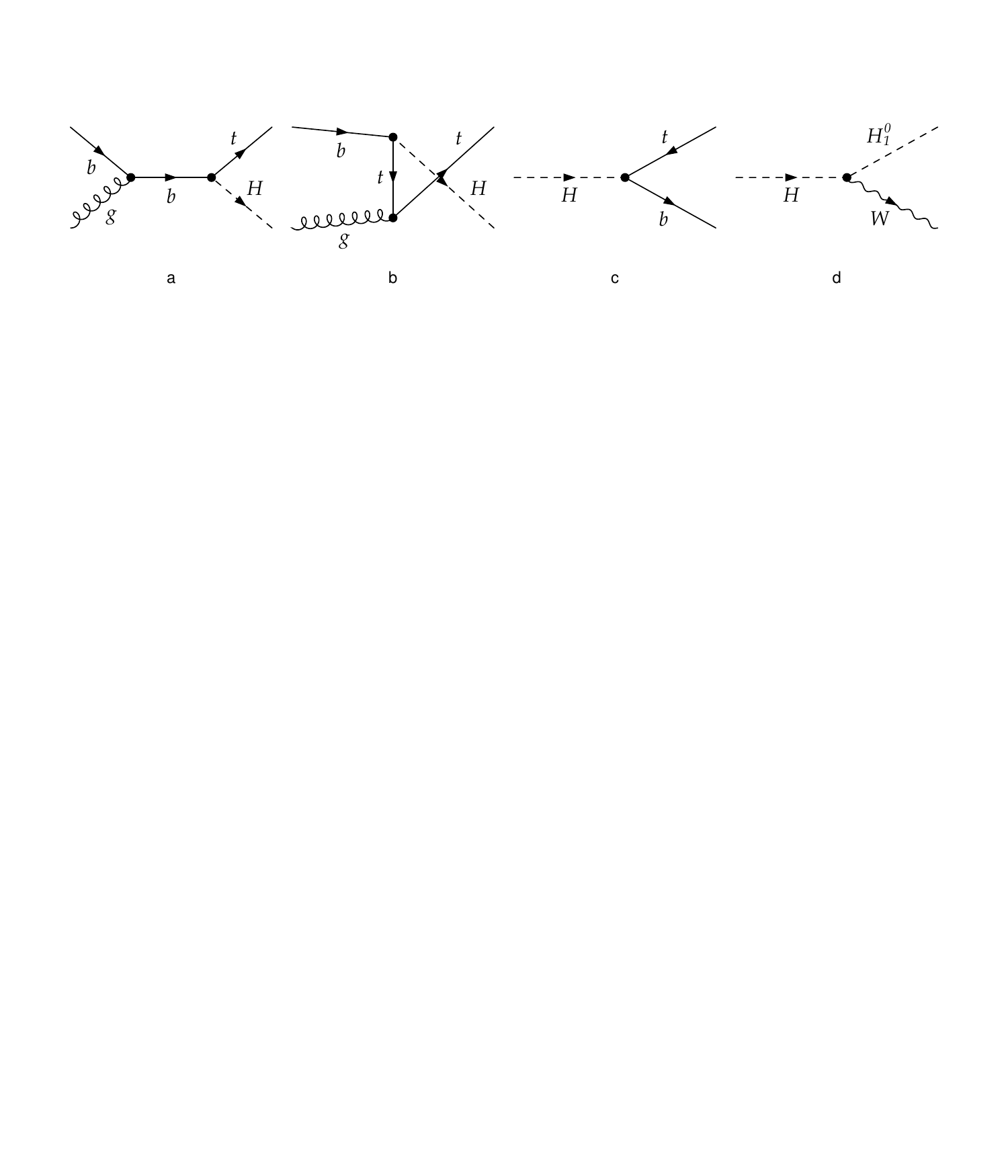}}}
 \vspace*{-13.5cm}
 \caption {The tree-level graphs for $b g \to t H^-$: a)~s-channel, b)~t-channel,
 and charged Higgs decays: c)~$H^- \to \bar{t}b$, d)~$H^- \to W^- H_1^0$.
  \label{tree}}}
  \end{center}
\end{figure}

We study CPV induced by one-loop corrections in the C2HDM in the following processes involving the charged Higgs boson:

$\bullet$ Decays:
\be
H^\pm \to t b\,,\qquad H^\pm \to W^\pm H_{i}\,, \qquad i=1,2\,.\label{decays2}
\ee
$\bullet$ Associated production with a top quark at the LHC:
\be
p p \to H^\pm t+X\,,\label{prod2}
\ee
with the partonic subprocesses:
\be
b~g \to t~H^- \quad {\rm and} \quad
\bar{b}~g\to \bar{t}~H^+ \label{procon}\,.~
\ee
$\bullet$ Charged Higgs production  (\ref{prod2}) plus subsequent decays
 (\ref{decays2}).\\
The tree-level graphs of the considered processes for $H^-$ are
shown in figure \ref{tree}.

\subsection{CP violating asymmetries}
\label{sec:CPasym}

For the processes listed in section~(\ref{sec:pro}) we investigate the following CP violating $H^\pm$ rate asymmetries:

 $\bullet$ Decay rate asymmetries $A_{D,f}^{CP}$, defined by:
\be
A_{D,f}^{CP}~(H^\pm\to f)=\frac{\Gamma (H^+\to f)-\Gamma (H^-\to \bar f)}
{2 \Gamma^{\rm tree} (H^+\to f) }\,,\label{CPVdecay}
\ee
where D stands for decay and $f$ stands for any of the decay modes: $f=t\bar{b};~W^\pm H_i^0$ with $i=1,2$.

 $\bullet$ Production rate asymmetry $A_P^{CP}$, related only to the process  (\ref{prod2}) and defined by:
\be
A_{P}^{CP}=\frac{\sigma (pp\to H^+\bar t)-\sigma
(pp\to H^-t)} {2 \sigma^{\rm tree} (pp\to H^+\bar t)}\,,\label{CPVprod}
\ee
where P stands for production.

 $\bullet$ Asymmetries $A^{CP}_{f}$ for the combined processes of production and subsequent decays, defined by:
\be
A^{CP}_{f}={\sigma(pp\rightarrow \bar{t}H^+\to  \bar{t}f)
-\sigma(pp\rightarrow t H^-\to t \bar f)\over
2 \sigma^{\rm tree}(pp\rightarrow \bar{t}H^+\to  \bar{t} f) }\,.
\label{AfCP}
\ee
In \cite{CPVinH+tMSSM} we have shown that in narrow width approximation, when the decay width of
$H^+$ is much smaller than its mass, the total asymmetry $A^{CP}_{f}$ given by
eq. (\ref{AfCP}) is an algebraic sum of the asymmetry $A_{P}^{CP}$ in the production and the asymmetry $A^{CP}_{D,f}$ in
the decay $f$ of the charged Higgs boson:
\be
A^{CP}_f= A_{P}^{CP}+A^{CP}_{D,f}\,. \label{finalf}
\ee

The decay asymmetries  (\ref{CPVdecay}) might be of interest for the ILC, where
CPV can occur only in the $H^\pm$ decays.
Since the measurements on $b\to s\gamma$ put a stringent lower limit
on the charged Higgs mass $M_{H^\pm}>295$ GeV, the decay modes  (\ref{decays2})
are dominant.

\subsection{CP violating loop contributions}

In order to get CPV using the asymmetries introduced in section \ref{sec:CPasym}, we need both
non-zero CP violating phases in the Lagrangian and CP conserving phases (strong phases) in the absorptive parts of the one-loop amplitudes.
\begin{figure}[h!]
\begin{center}
\vspace*{-2cm}
{\mbox{\resizebox{!}{18.0cm}{\includegraphics{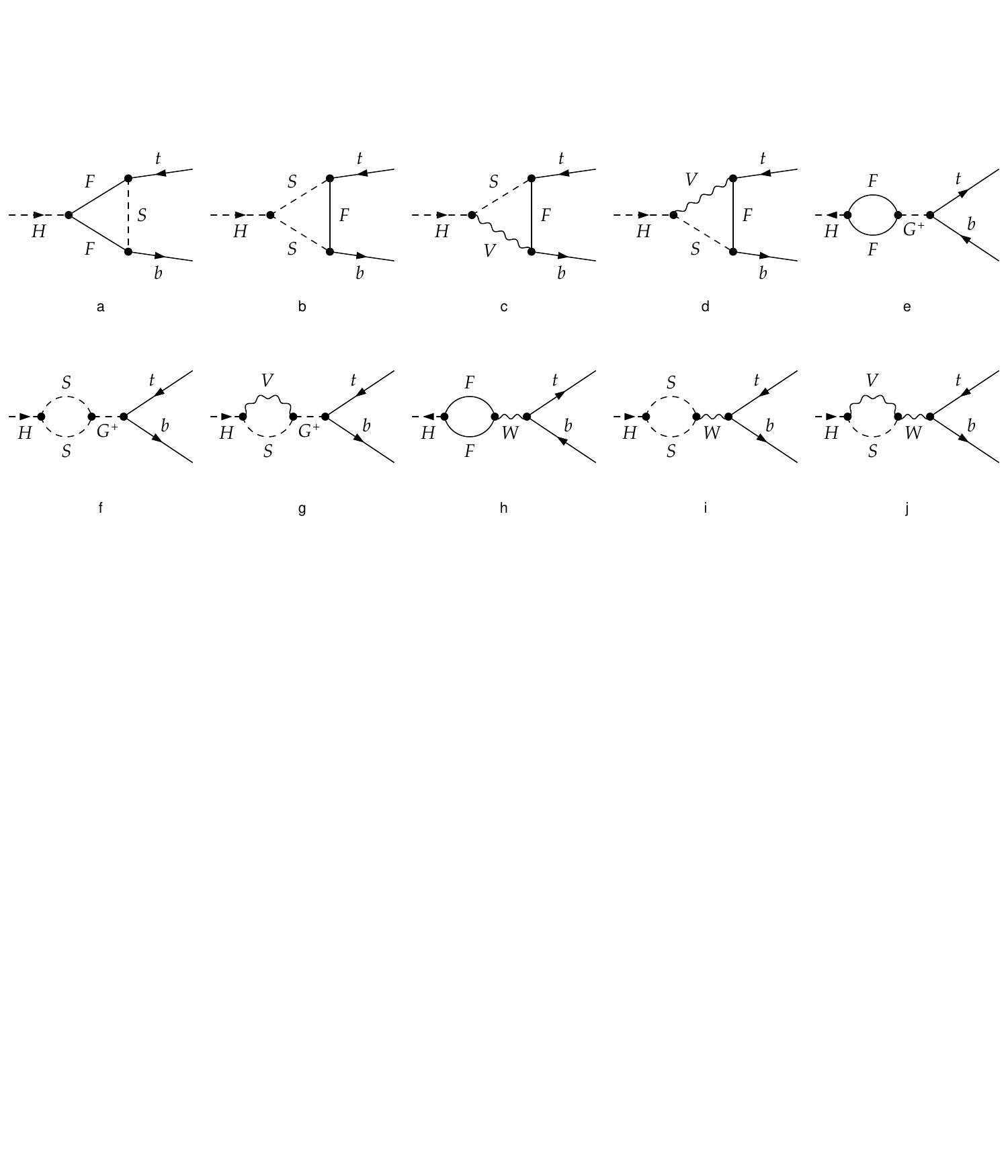}}}
\vspace*{-9.5cm} \caption{Generic CP violating selfenergy and vertex
contributions to $H^- \to \bar{t}b$. F denotes a generic
fermion, S a generic neutral or charged scalar field,
and V a generic vector boson. \label{hdecay}}}
  \end{center}
\end{figure}

In the C2HDM, the CP violating phases arise from:
\begin{itemize}
\item Neutral Higgs couplings to a fermion pair
\item Charged Higgs -- Neutral Higgs -- gauge bosons couplings
\item Charged Higgs -- Neutral Higgs -- Goldstone bosons couplings
\end{itemize}
The parts of the C2HDM Lagragian
needed for this study are given in the appendix \ref{App:Lag}.

The CP conserving phases originate
from various on-shell intermediate states of the one-loop amplitudes.
For the considered $H^\pm$ production and decay processes,
the strong phases coming from cuts e.g. in $t\to bW^\pm$ and $H_1^0\to b\bar{b}$ will always contribute,
while the strong phases coming from cuts in
$H^\pm \to W^\mp H_i^0$, $H^\pm \to G^\mp H_i^0$,
$H^\pm \to t\bar{b}$, etc. in the loops,
 will contribute only if they are kinematically allowed.
 %
\begin{figure}[h!]
\begin{center}
\vspace*{-1cm}
{\mbox{\resizebox{!}{18.0cm}{\includegraphics{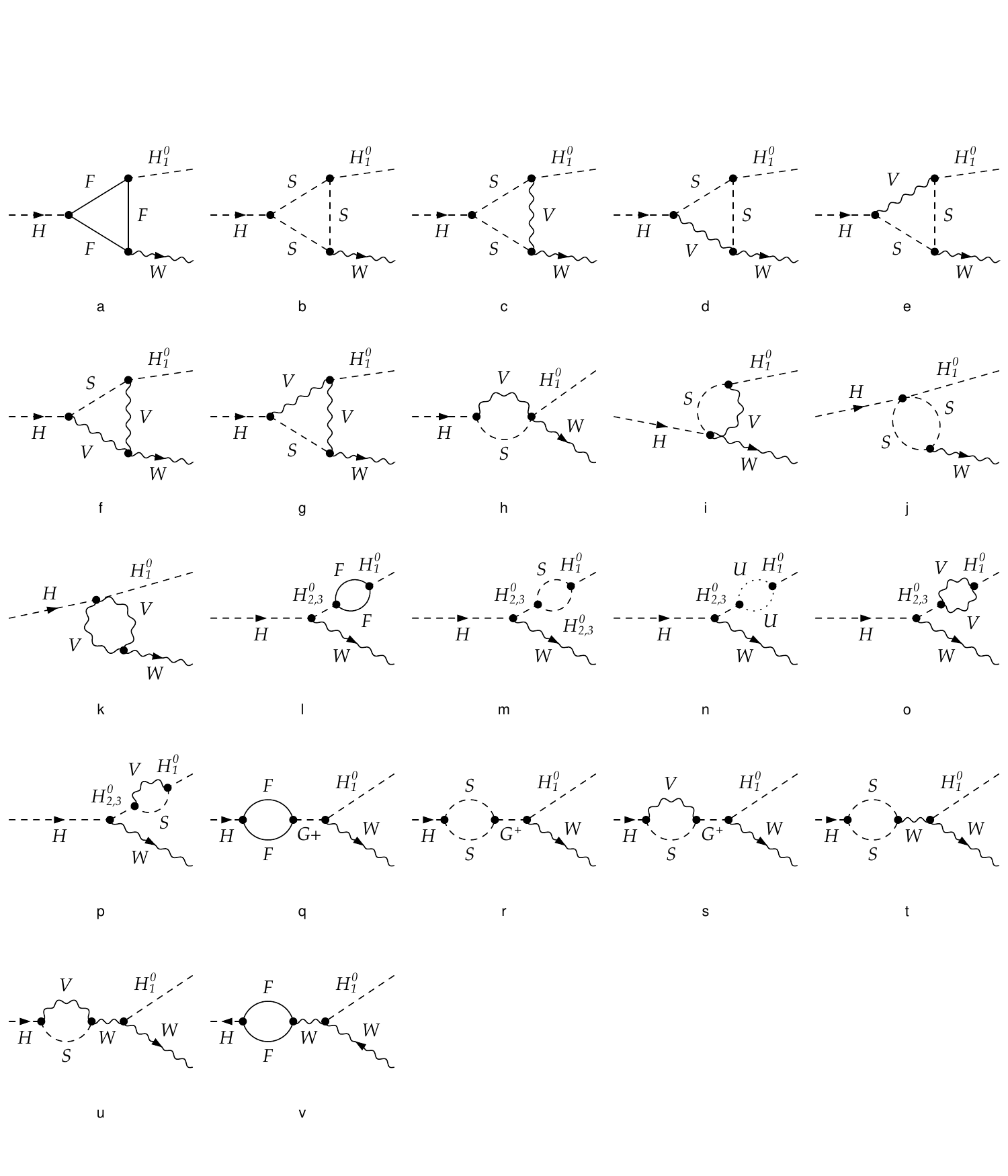}}}
\caption{Generic CP violating selfenergy and vertex contributions
to $H^- \to W^- H_1^0$. \label{hdecaywh}}}
  \end{center}
\end{figure}
All possible generic CP violating one-loop contributions to the decay $H^\pm \to t\bar{b}$ are shown in
figure~\ref{hdecay}
and to the decay $H^\pm \to W^\pm H_1^0$  in figure~\ref{hdecaywh}.

\begin{figure}[h!]
\begin{center}
\vspace*{-8cm}
{\mbox{\resizebox{!}{18.0cm}{\includegraphics{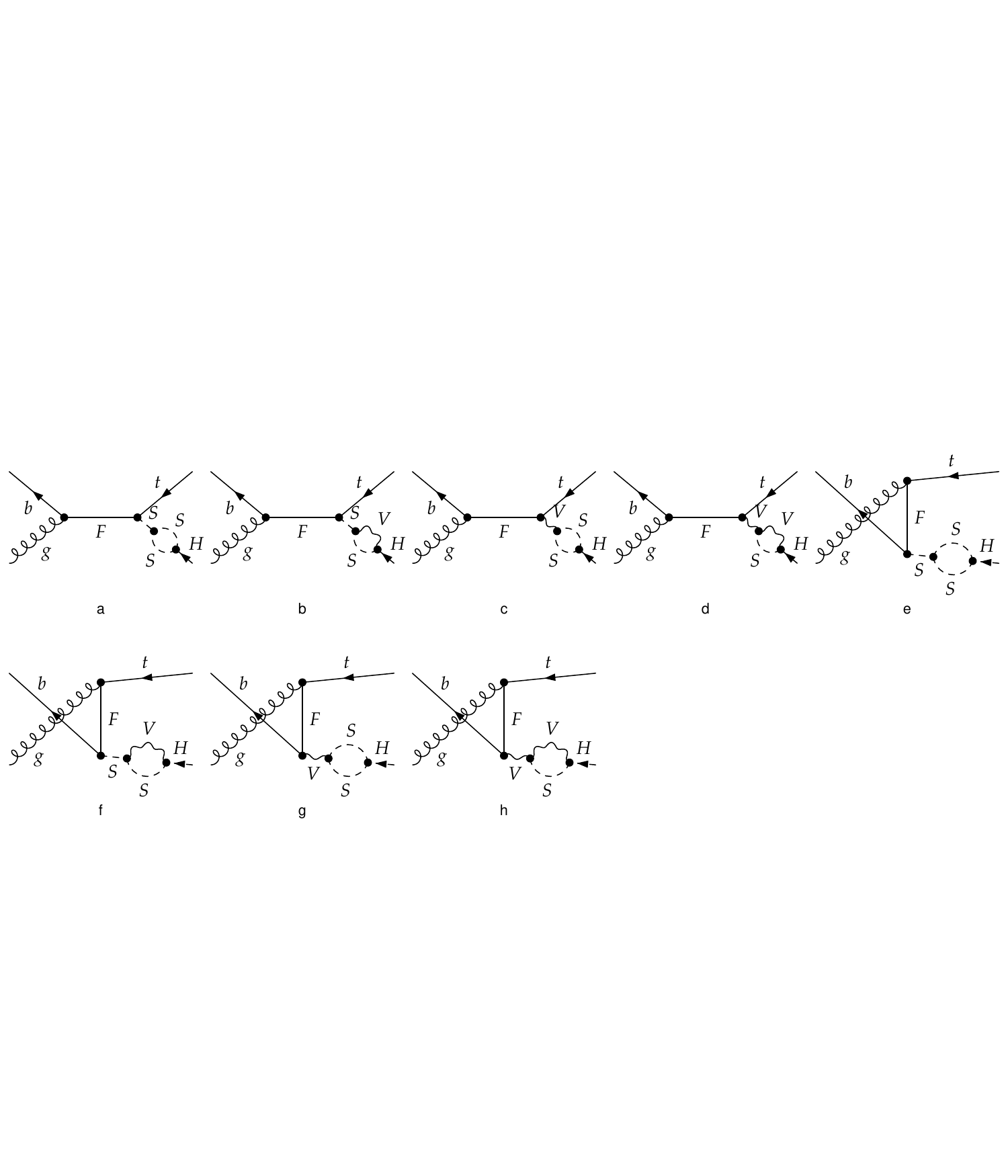}}}
\vspace*{-5.5cm} \caption{Generic CP violating selfenergy
contributions to charged Higgs boson production $\bar{b}~g \to
H^+ \bar{t}$. \label{selfes}}}
  \end{center}
\end{figure}

\begin{figure}[h!]
\begin{center}
\vspace*{-8cm}
{\mbox{\resizebox{!}{18.0cm}{\includegraphics{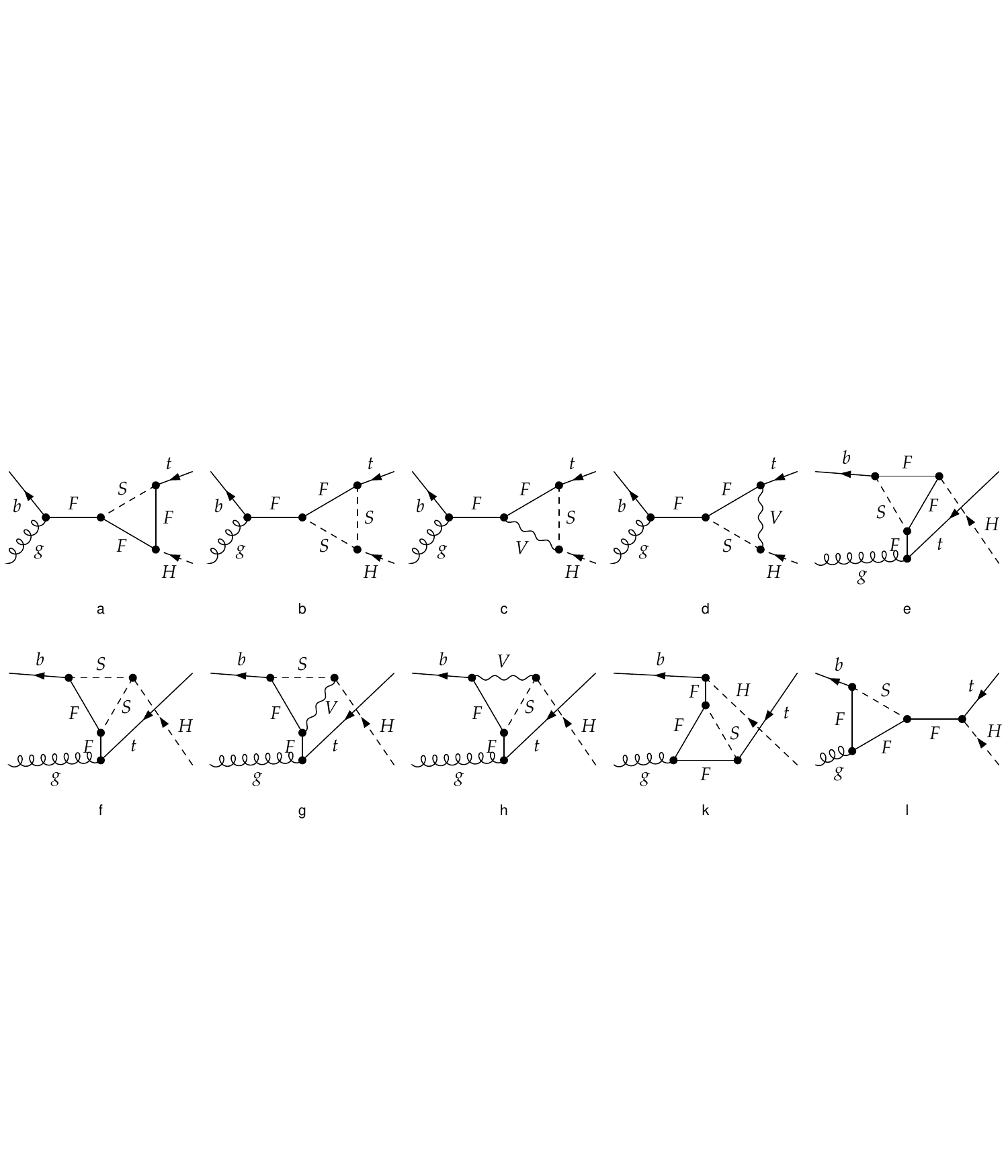}}}
\vspace*{-5.5cm}
  \caption {Generic CP violating vertex contributions to charged Higgs boson production $\bar{b}~g \to
H^+ \bar{t}$. \label{vertex}}}
  \end{center}
\end{figure}

\begin{figure}[h!]
\begin{center}
\vspace*{-7cm}
{\mbox{\resizebox{!}{18.0cm}{\includegraphics{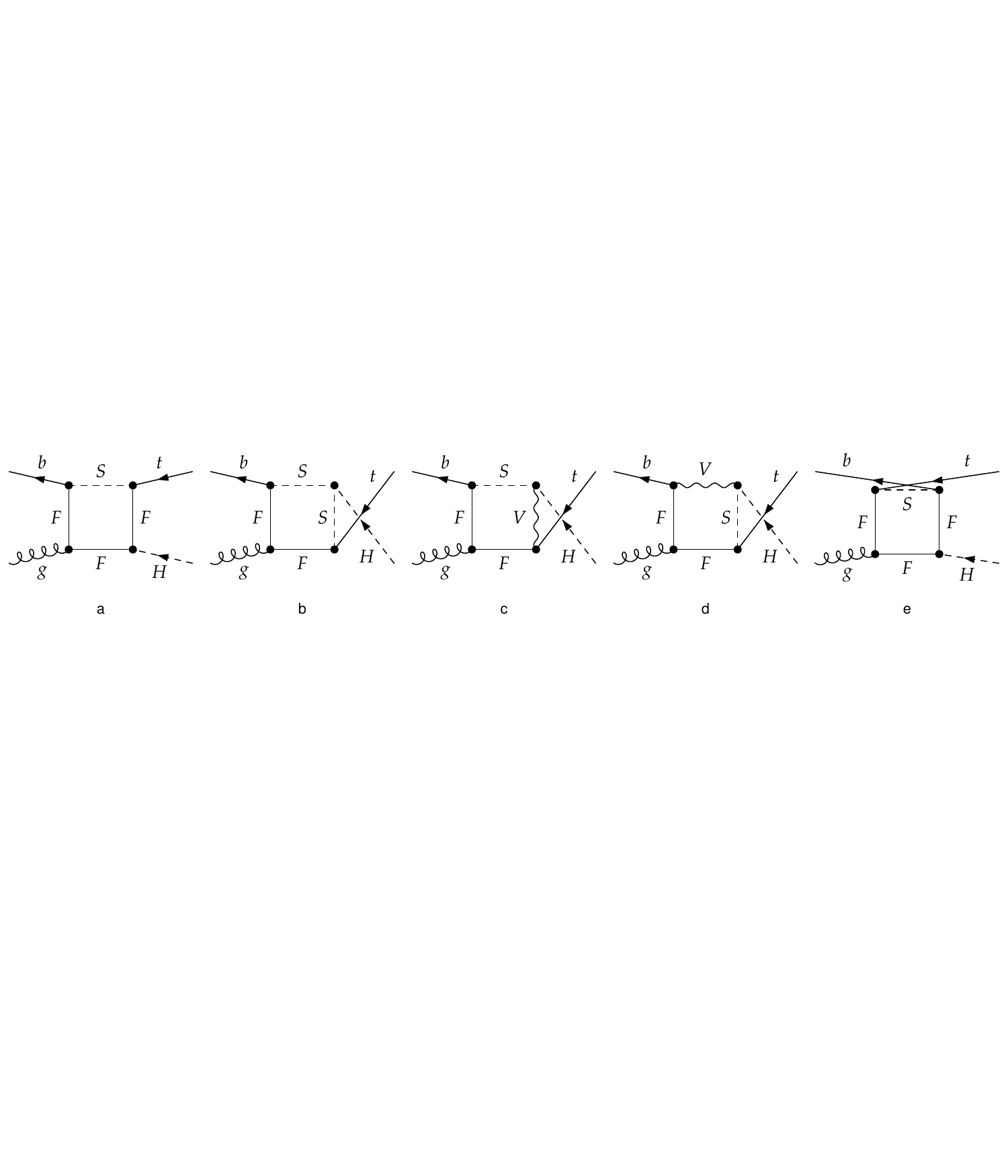}}}
\vspace*{-8.7cm}
  \caption {Generic CP violating box contributions to charged Higgs boson production $\bar{b}~g \to
H^+ \bar{t}$. \label{boxes}}}
\end{center}
\end{figure}

Furthermore, we have three types of possible generic CP violating
loop contributions to the partonic cross sections of the production processes  (\ref{prod}):
selfenergy contributions, shown in figure \ref{selfes}; vertex contributions shown in figure \ref{vertex} and
box diagram contributions shown in figure \ref{boxes}.
All diagrams have been generated with
FA \cite{FeynArts} package, for which we have
written a complete model file for the C2HDM, see Appendices \ref{App:feynarts} and \ref{App:formcalc} for details.

\section{Numerical results}
\label{sec:numerics}

In this section we present our numerical results for the CP
violating asymmetries (\ref{CPVdecay}), (\ref{CPVprod}) and
(\ref{AfCP}) in the C2HDM with a softly broken $Z_2$ symmetry
 (\ref{THDMpot}). All calculations have been done using the
packages FA and FC \cite{FeynArts}, for which we have written a
complete FA model file and have extended the corresponding FC fortran
drivers for the C2HDM. The implementation is described for the FA model file in
the appendix \ref{App:feynarts}  and for the FC fortran drivers in
\ref{App:formcalc}. In our numerical
analysis we also have used LoopTools \cite{LT,LT2,LT3}. The calculations
are done in the 'tHooft-Feynman gauge using dimensional
regularization. However, we stress that we do not renormalize any
parameters or fields since our CPV rate asymmetries involve only the imaginary
parts of the loop integrals which are always finite. For the evaluation
of the PDF's we use CTEQ5L, with $\alpha_s$,
calculated at the scale $Q=\sqrt{\hat{s}}$.

In this paper we  work with the following set of real input parameters \cite{ElKaffas}, see section \ref{sec:Z2param}:
\be
\bigg\{M_{H_1^0},~~ M_{H_2^0},~~ M_{H^+},~~ \alpha_1,~~
\alpha_2,~~ \alpha_3,~~\tan \beta,~~ {\rm Re}\, (m_{12})
\bigg\}\,.
\ee
In the literature often the parameter $\mu$ is used instead of ${\rm Re}\,( m_{12})$:
\be
\mu^2 = \frac{v^2}{2v_1v_2}\,{\rm Re}\, ( m_{12}^2)\,, \quad v^2=v_1^2+v_2^2\,.\label{mu}
\ee
The expressions for the parameters of the scalar potential  (\ref{THDMpot}), $\l_i$, $i=1,2,3,4,5$,
as functions of the physical masses, mixing angles and $\mu$ \cite{theor.constr.} are given in
appendix \ref{App:lambda}.
For the values of the SM parameters used we refer to
\cite{Amsler:2008zzb}, except for the
top mass, taken from the last Tevatron measurement \cite{Tev:2009ec}:
\be
m_t=173.1\, {\rm GeV}\,,\quad m_b=4.7\,  {\rm GeV}\,,\quad {\rm and} \quad
\alpha=1/137.03599.
\ee

Furthermore, we would like to add some comments on the existing
constraints on the values of the C2HDM parameters.
Basically they come from:

$\bullet$ theory -- these are the requirements for positivity and
unitarity of the Higgs potential, see section \ref{sec:th.constr.}

$\bullet$ experiment -- mainly coming from
the electroweak precision data at LEP. These constraints we will discuss in the next subsection

\subsection{Experimental constraints}
\label{sec:exp.constr.}

In principle, there is quite a long list of experimental bounds to constrain the
phenomenology of the 2HDM. Examples are: B-mixing/B-decays
constraints, LEP2 non-discovery constraints, direct searches for
$H^\pm$, muon anomalous magnetic moment, electron electric dipole
moment (EDM), etc., see \cite{ElKaffas, Per2, exp.constr., Per3} and the references therein. Many of these constraints
are strongly dependent on the model of Yukawa interactions, on the
considered process and on the related parameter space.
As already mentioned, the theoretical constraints described in section \ref{sec:th.constr.} already
strongly reduce the parameter space of the C2HDM \cite{Per3}.
In \cite{WahabElKaffas:2007xd} the authors combine these constraints with the existing experimental constraints and
review the  "profile of the surviving parameter space" of the
type II 2HDM. There they show that large values of $\tan \beta$ are forbidden by the unitarity constraints, see section \ref{sec:th.constr.},
except for the case $M_{H^0_1} < \mu$ \cite{WahabElKaffas:2007xd}. Furthermore, they show that
the experimental constraints also usually exclude parameter regions for $\tan \beta \sim 10$ \cite{WahabElKaffas:2007xd}.
However, in principle it can be shown that one can fine-tune the parameters to allow some tiny parameter regions for large $\tan \beta$ \cite{largetanb}.

Among the experimental constraints we consider the following ones, which have the strongest impact on
restricting the general 2HDM parameter space:

$\bullet$ For the lightest neutral Higgs boson we take into account the LEP2 non-discovery bound, $M_{H_1^0} > 114.4$ GeV \cite{pdg}.

$\bullet$ In addition to the lower bounds from the CERN LEP and Tevatron
direct searches \cite{Ref:LEP,Ref:LEP2,Ref:Tevatron,Ref:Tevatron2}, the charged Higgs
mass is quite severely constrained by the $B\to X_s\gamma$ data
 \cite{Ref:bsgNLO,Ref:bsgNLO2,Ref:bsgNLO22,Ref:bsgNLO23,Ref:bsgNNLO,Ref:bsgNNLO_THDM,Ref:bsgNNLO_THDM2}. At
the next-to-next-to-leading order in the type-II 2HDM it implies
that $M_{H^\pm}\geq 295$ GeV,
 \cite{Ref:bsgNNLO,Ref:bsgNNLO_THDM,Ref:bsgNNLO_THDM2}. In our analysis we follow
the latter lower bound.

$\bullet$ Another important experimental constraint, coming from the
electroweak physics, is related to the precise determination of the
$\rho$-parameter \cite{Ross}. Its deviation $\Delta\rho$ from the SM value
should accommodate all new physics contributions. $\Delta \rho$ is constrained
by the error of the measured  value of the parameter $\rho_0$, which at the 2 $\sigma$
level is \cite{pdg}:
\be
\rho_0=1.0004^{~+ 0.0029}_{~- 0.0011}~.
\ee
In our numerical code we implement the expressions for $\Delta
\rho$ given in \cite{ElKaffas},
with the requirement:
\be
- 0.0011\le\Delta\rho\le0.0029\,.
\ee
The analytic expressions for the extra contributions in
$\Delta\rho$ in the framework of the C2HDM are given in appendix \ref{App:rho}.

\subsection{CP violation in $H^\pm$-decays}\label{sec:numdecays}

We discuss the CPV decay rate asymmetries $A_{D,f}^{CP}$, given with eq. (\ref{CPVdecay}),
for the decays $H^\pm \to t\bar{b}$ and $H^\pm \to W^\pm H_i^0,\, i=1,2$.
Note that we do not consider the decay mode $H^\pm \to W^\pm H_3^0$
as a possible measurable channel for the CPV asymmetry due to its tiny branching ratio
(BR) in the considered parameter ranges.
\begin{figure}[!ht]
\begin{center}
\begin{tabular}{cc}\hspace*{-1cm}
\resizebox{85mm}{!}{\includegraphics{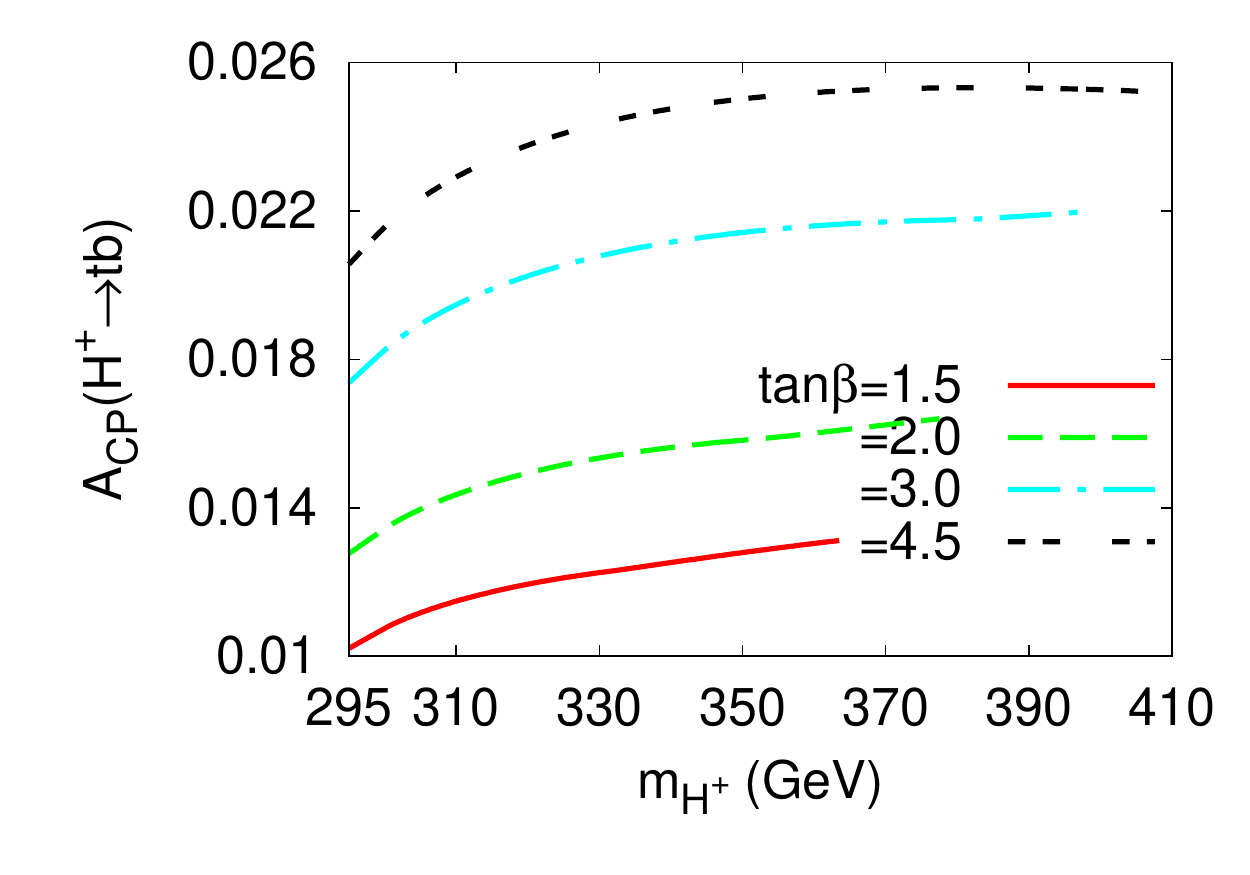}} & \hspace{-1.cm}
\resizebox{85mm}{!}{\includegraphics{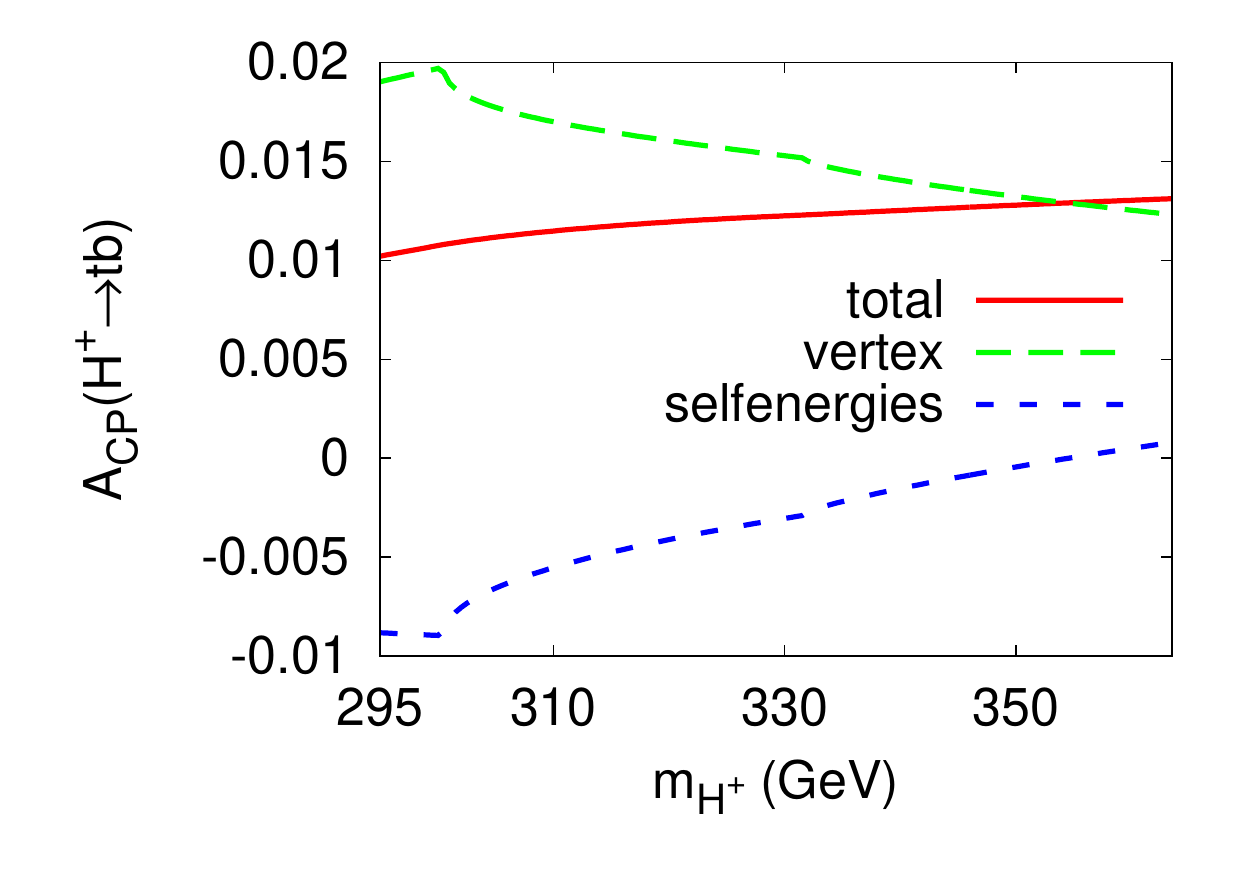}}
\end{tabular}
\end{center}
\caption{Left: CPV asymmetry $A_{D,tb}^{CP}$
as a function of $M_{H^+}$ for four values of $\tan\beta$.
The other parameters are: $M_{H_1^0}= 120$ GeV, $M_{H_2^0}= 220$ GeV,
${\rm Re} (m_{12})=170$ GeV, $\alpha_1=0.8$, $\alpha_2=-0.9$ and $\alpha_3=\pi/3$. Right: CPV vertex
and selfenergy contributions to $A_{D,tb}^{CP}$,
as functions of the charged Higgs mass for $\tan\beta=1.5$.}
\label{Hptotb}
\end{figure}
\begin{figure}[!ht]
\begin{tabular}{cc}\hspace*{-1cm}
\resizebox{85mm}{!}{\includegraphics{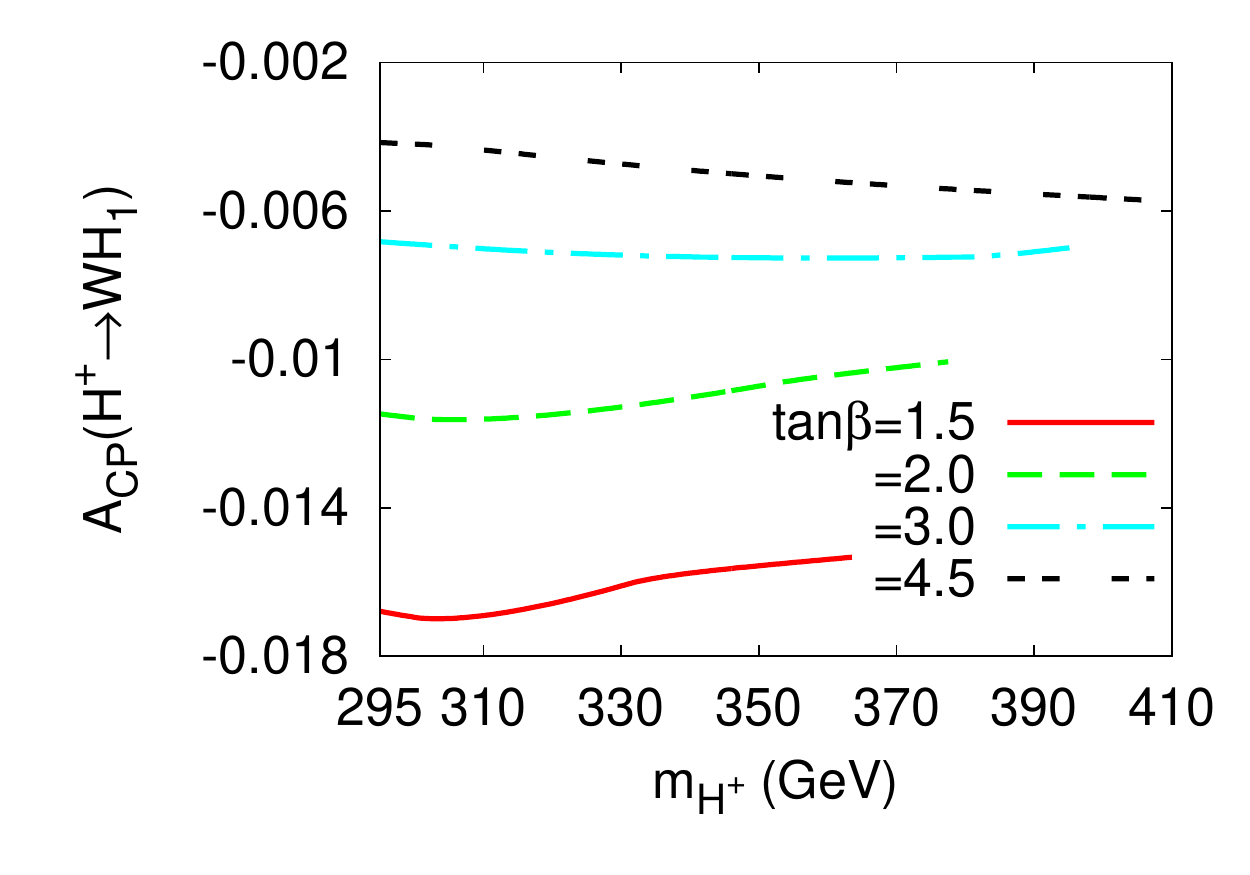}} & \hspace{-1.cm}
\resizebox{85mm}{!}{\includegraphics{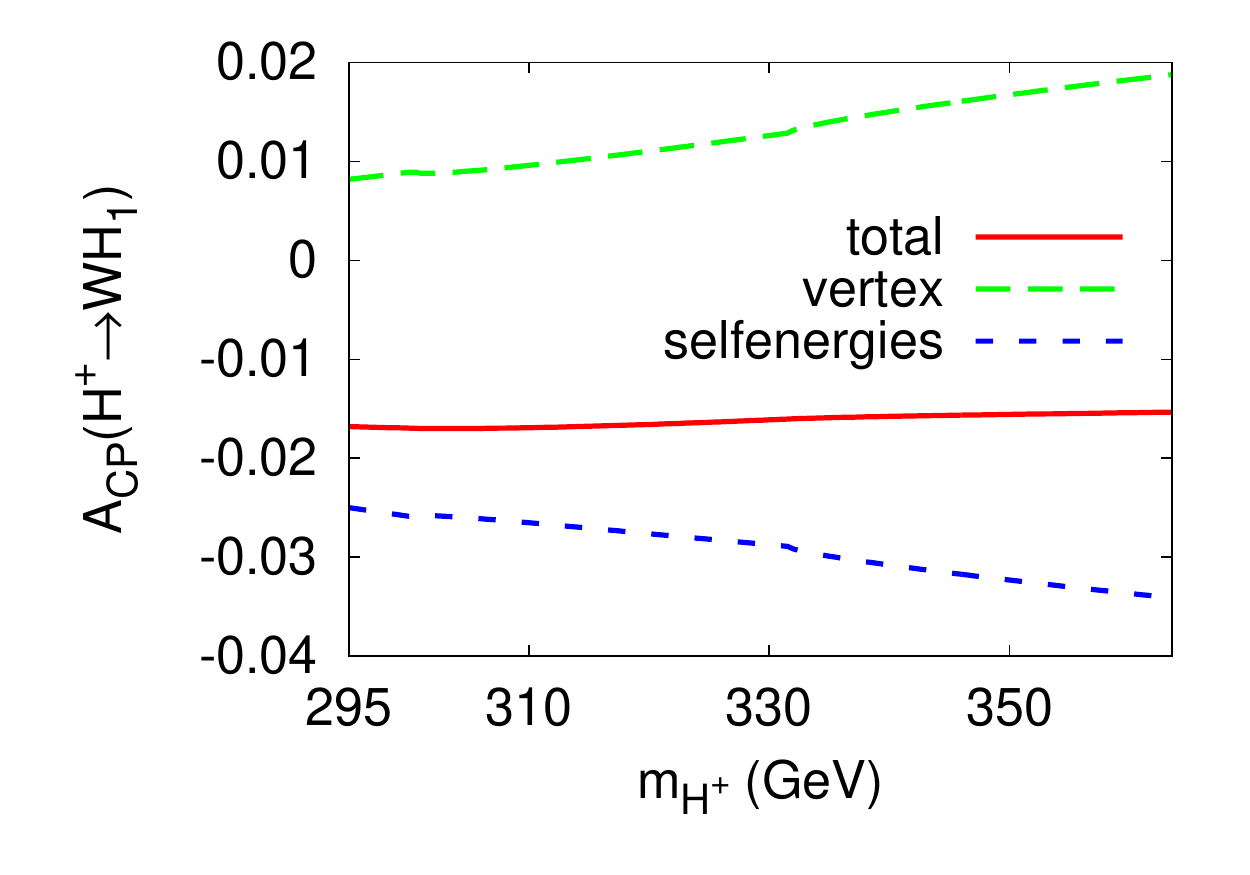}}
\end{tabular}
\caption{Left: CPV asymmetry $A_{D,WH_1^0}^{CP}$
as a function of $M_{H^+}$ for four values of $\tan\beta$.
The other parameters are the same as for figure \ref{Hptotb}. Right: CPV vertex
and selfenergy contributions to $A_{D,WH_1^0}^{CP}$,
as functions of the charged Higgs mass for $\tan\beta=1.5$.}
\label{decaycancel}
\end{figure}
\begin{figure}[!ht]
\begin{center}
\begin{tabular}{cc}\hspace*{-1cm}
\resizebox{85mm}{!}{\includegraphics{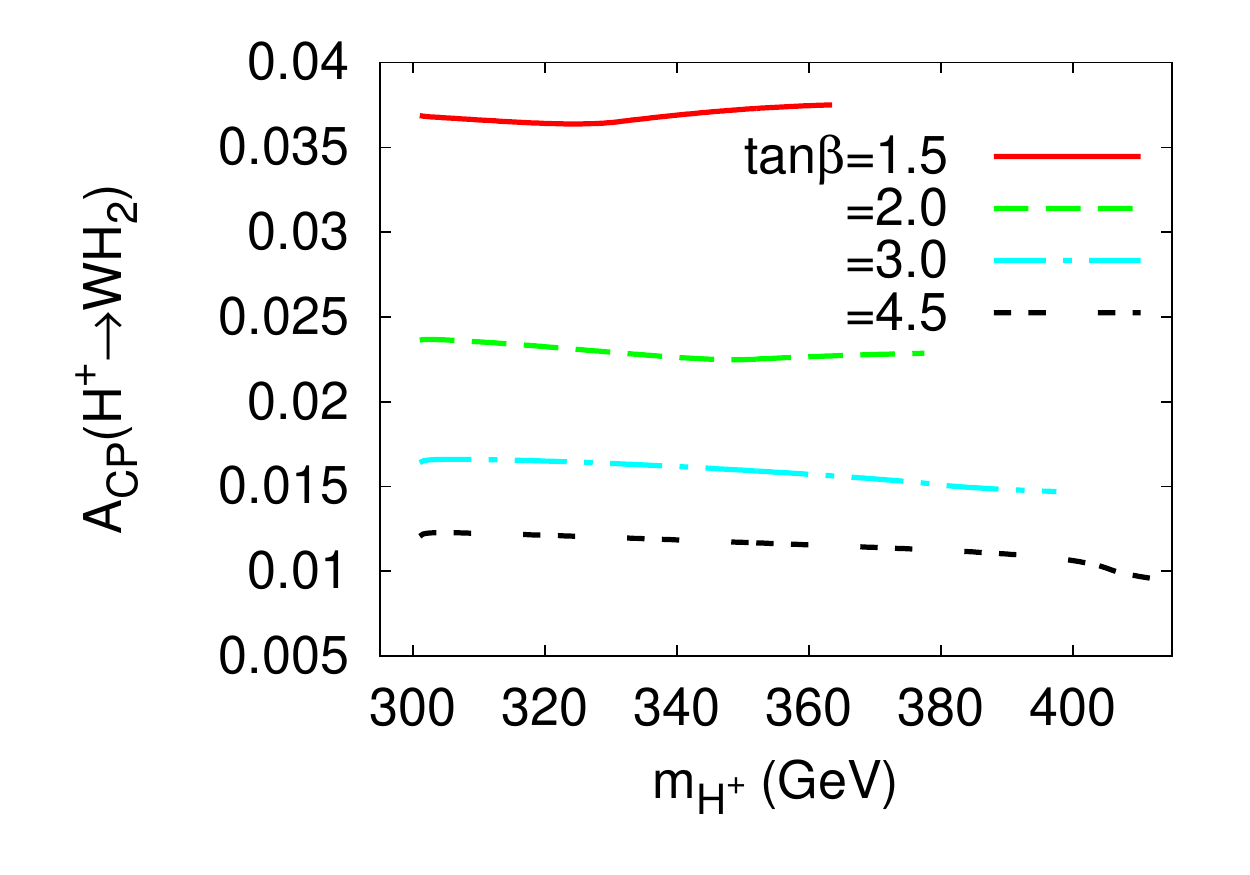}}
\end{tabular}
\end{center}
\caption{CPV asymmetry $A_{D,WH_2^0}^{CP}$  (\ref{CPVdecay}),
as a function of $M_{H^+}$ for four values of $\tan\beta$.
The other parameters are: $M_{H_1^0}= 120$ GeV, $M_{H_2^0}= 220$ GeV,
${\rm Re} (m_{12})=170$ GeV, $\alpha_1=0.8$, $\alpha_2=-0.9$ and $\alpha_3=\pi/3$.}\label{HptoWH}
\end{figure}

In figure~\ref{Hptotb}, figure~\ref{decaycancel} and figure~\ref{HptoWH} we present the CP
asymmetries $A_{D,tb}^{CP}$~, $A_{D,WH_1^0}^{CP}$ and $A_{D,WH_2^0}^{CP}$, respectively,
as functions of the charged Higgs mass, for several values of $\tan\beta$.
The curves in the figures are cut once any of the
parameter constraints discussed in section \ref{sec:th.constr.} and section \ref{sec:exp.constr.} is violated.
In particular, in the considered parameter range, the $\Delta\rho$ constraint
imposes that the charged Higgs mass $M_{H^+}$ cannot acquire large values:
in figure \ref{Hptotb}, for $\tan\beta = 1.5$, $M_{H^+}$ cannot
exceed the limit of $\approx 368$ GeV,
while for $\tan\beta = 4.5$: $M_{H^+}\lsim 408$ GeV, etc..

All discussed CPV asymmetries exhibit a mild dependence on $M_{H^+}$ and a strong dependence on $\tan \beta$:
 $A_{D,tb}^{CP}$ increases, while $A_{D,WH_1^0}^{CP}$ and $A_{D,WH_2^0}^{CP}$ strongly decrease
when $\tan \beta$ increases. This is shown in figures \ref{Hptotb}, \ref{decaycancel}, and \ref{HptoWH}.

As we already commented in section \ref{sec:exp.constr.} large values of $\tan\beta$ are
excluded, except for some small areas in the region $M_{H^0_1} < \mu$ which require fine-tuning of the parameters
\cite{WahabElKaffas:2007xd}. Moreover, we have checked that
 very often for large $\tan\beta$ the mass of the heaviest Higgs boson $M_{H_3^0}$
becomes tachyonic. Therefore in most of the cases we show numerical results for $\tan\beta$ in the range $1.5 \div 4.5$.

In figure \ref{Hptotb} we see that the asymmetry $A_{D,tb}^{CP}$ is positive
and can reach $\sim$ 2.5\% for $\tan\beta=4.5$ and $M_{H^+}\approx 400$ GeV. In the same figure we show the
vertex and the selfenergy contributions in $A_{D,tb}^{CP}$ for $\tan \beta=1.5$.
The figure shows that these contributions enter with opposite signs and there is a partial
cancellation between them. In contrast to the MSSM, here the dominant
contribution in the CPV asymmetry comes from the vertex diagrams.

For the bosonic channel $H^\pm \to W^\pm H_1^0$, the CPV asymmetry is shown in figure \ref{decaycancel}.
It is seen that $A^{CP}_{D, WH_1^0}$ is negative
and can reach $\sim$ 1.7\% for $\tan \beta =1.5$ and $M_{H^+}$ close to the lower limit $\approx 297$ GeV.
In this channel there is also a cancellation between
the selfenergy and the vertex contributions, which is shown in the figure. In the CPV decay
asymmetry $A^{CP}_{D, WH_1^0}$ the dominant contribution stems from the selfenergies.

In figure \ref{HptoWH} we show the asymmetry $A^{CP}_{D, WH_2^0}$ as a function of the charged Higgs mass. In contrast to
the other bosonic channel $H^\pm \to WH_1^0$, it is positive and the absolute value is larger. For $\tan \beta=1.5$, it is almost
constant $\sim$3.7\%.
For this asymmetry we do not show explicitly the individual contributions. However, we would like to note, that
cancellation occurs between the vertex and the selfenergy contributions. Similar to the decay into $H_1^0$
the selfenergy contributions are dominant in $A^{CP}_{D, WH_2^0}$.

In all three discussed cases we have checked that the influence of the
CKM matrix is very small, therefore we work with a diagonal one.

\begin{figure}[!ht]
\centering
\includegraphics[height=3.0in,width=7cm]{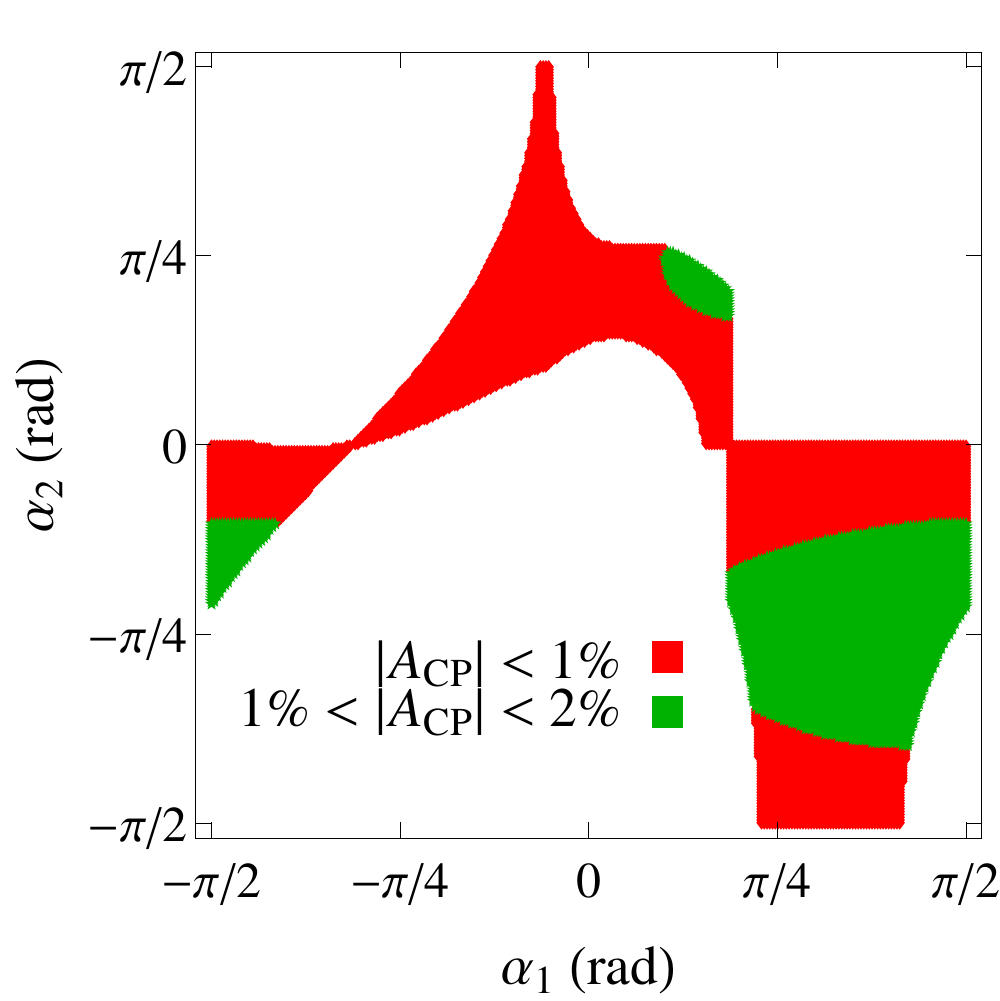}
\hskip 0.1cm
\includegraphics[height=3.0in,width=7cm]{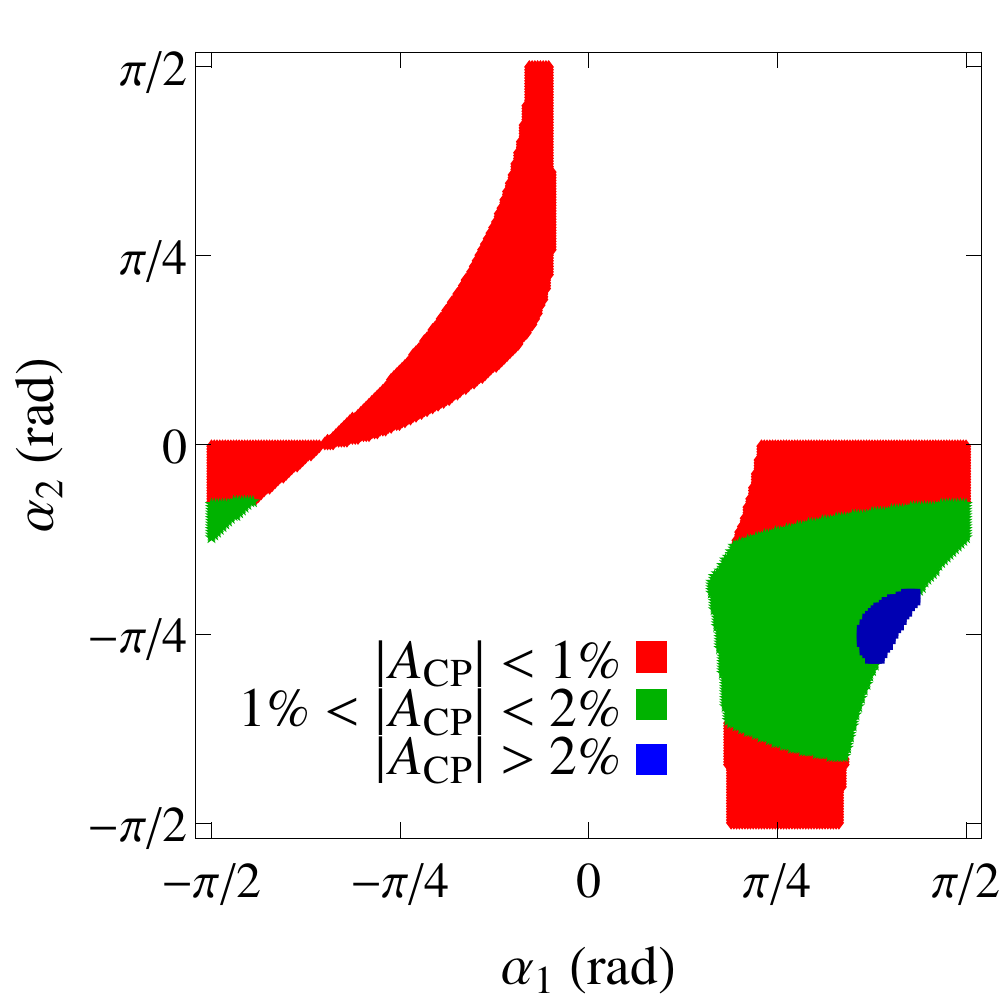} \\
\includegraphics[height=3.0in,width=7cm]{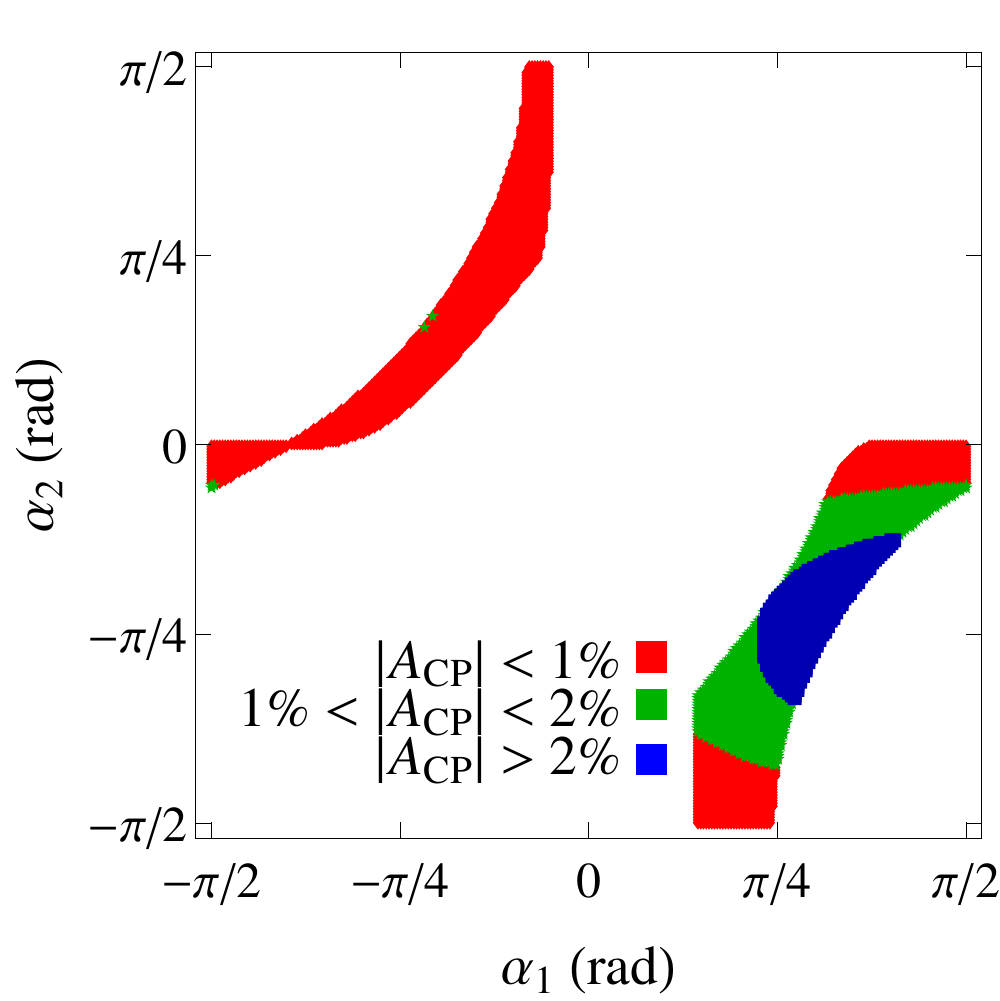}
\hskip 0.1cm
\includegraphics[height=3.0in,width=7cm]{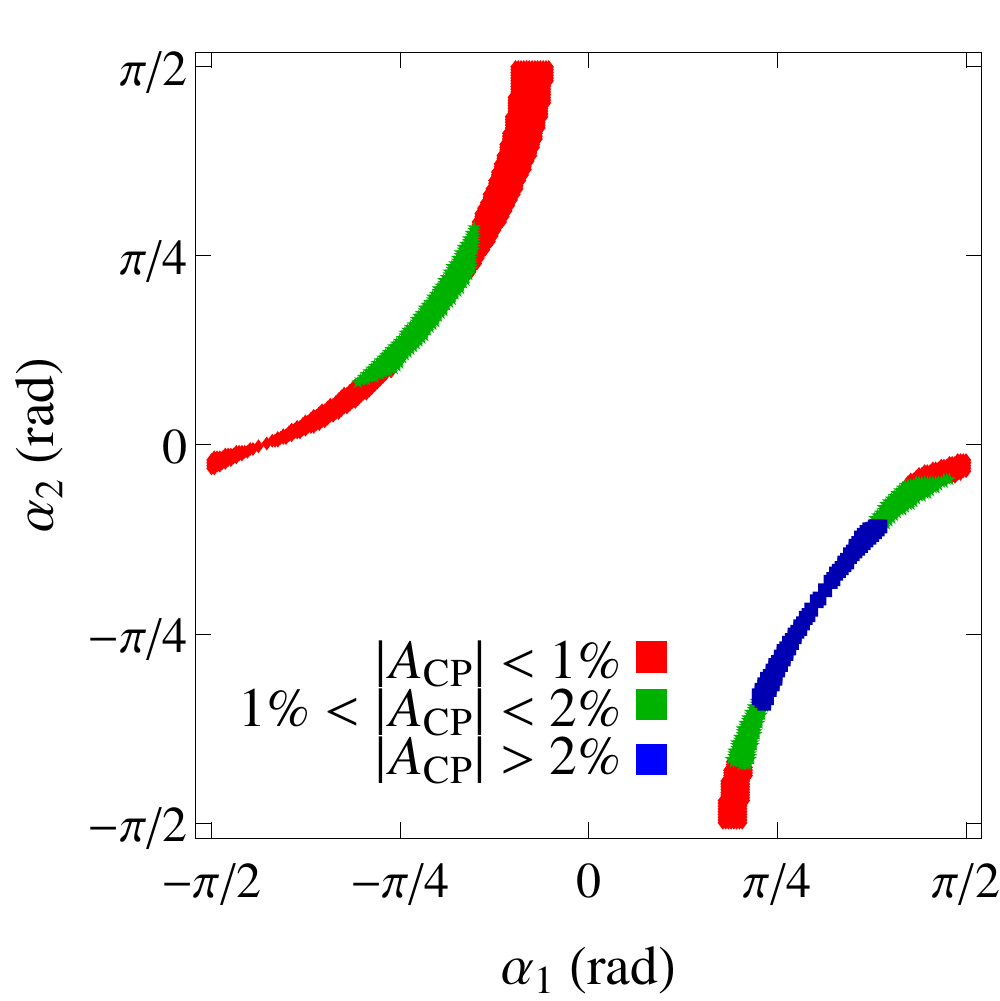}
\caption{The allowed parameter regions in the $(\alpha_1,\alpha_2)$ plan
in the C2HDM together with the absolute value of the CPV asymmetry $A_{D,tb}^{CP}$.
We have taken $M_{H_1^0}= 120$ GeV, $M_{H_2^0}= 220$ GeV,
$M_{H^\pm} = 350$ GeV, ${\rm Re} (m_{12})=170$ GeV, and $\alpha_3=\pi/3$.
On the top left plot $\tan\beta=1.5$,
top right: $\tan\beta=2$, down left: $\tan\beta=3$,
and down right: $\tan\beta=4.5$. }
\label{scan1}
\end{figure}

In figure \ref{scan1} we show a scan over
the parameters $\alpha_1$ and $\alpha_2$, in order to illustrate the allowed domain for $(\alpha_1,\alpha_2)$ together with
the size of the CPV asymmetry in the $H^\pm \to t\bar{b}$ decay. As we already mentioned,
the excluded parameter regions are due to the combination of all theoretical and experimental constraints:
vacuum stability, unitarity, large additional contributions to $\Delta\rho$, as well as
tachyonic modes of $M_{H_3^0}$ or violation of the ordering
$M_{H_1^0}\leq M_{H_2^0} \leq M_{H_3^0}$.
In the figure we see that for smaller values of $\tan\beta$ the allowed domain is larger and
almost vanishes for $\tan\beta=4.5$ and that
the CPV asymmetry $A^{CP}_{D,tb}$ exceeds 2\% only in some small areas.
\begin{figure}[!ht]
\centering
\includegraphics[height=3.0in,width=7cm]{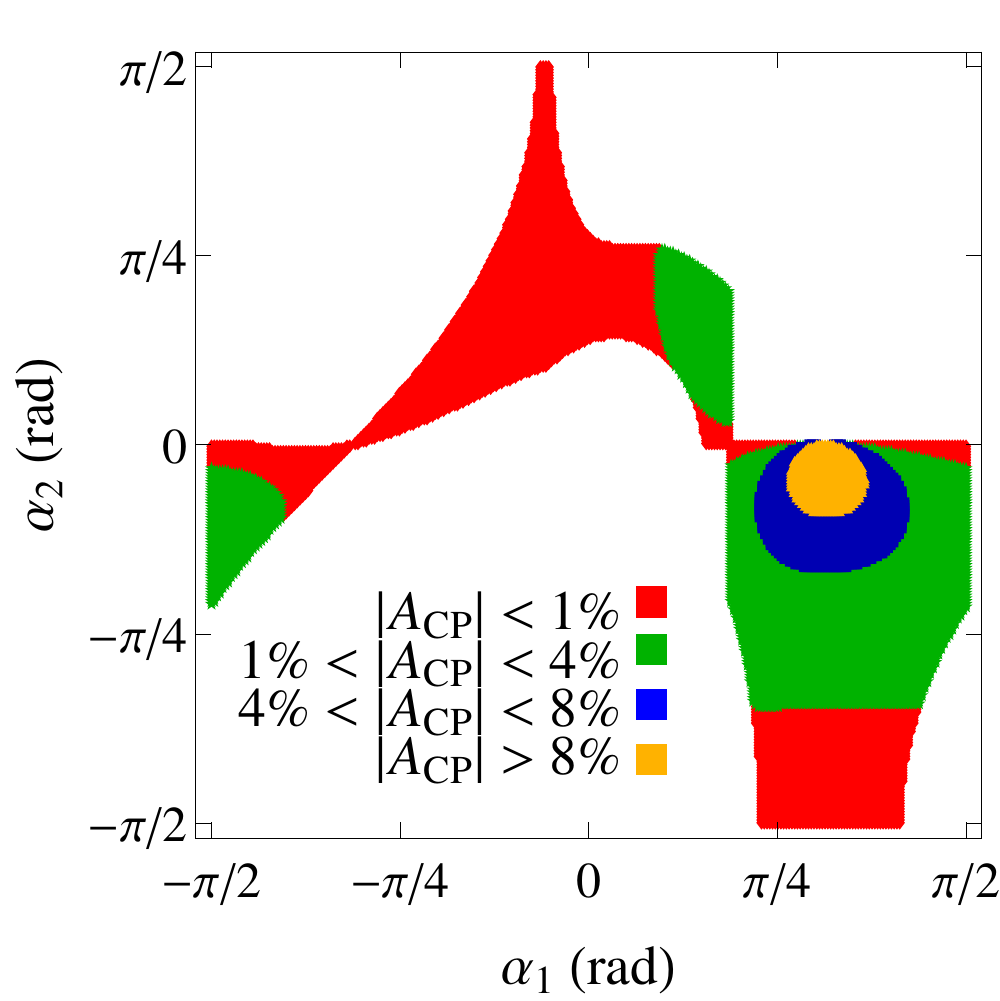}
\hskip 0.1cm
\includegraphics[height=3.0in,width=7cm]{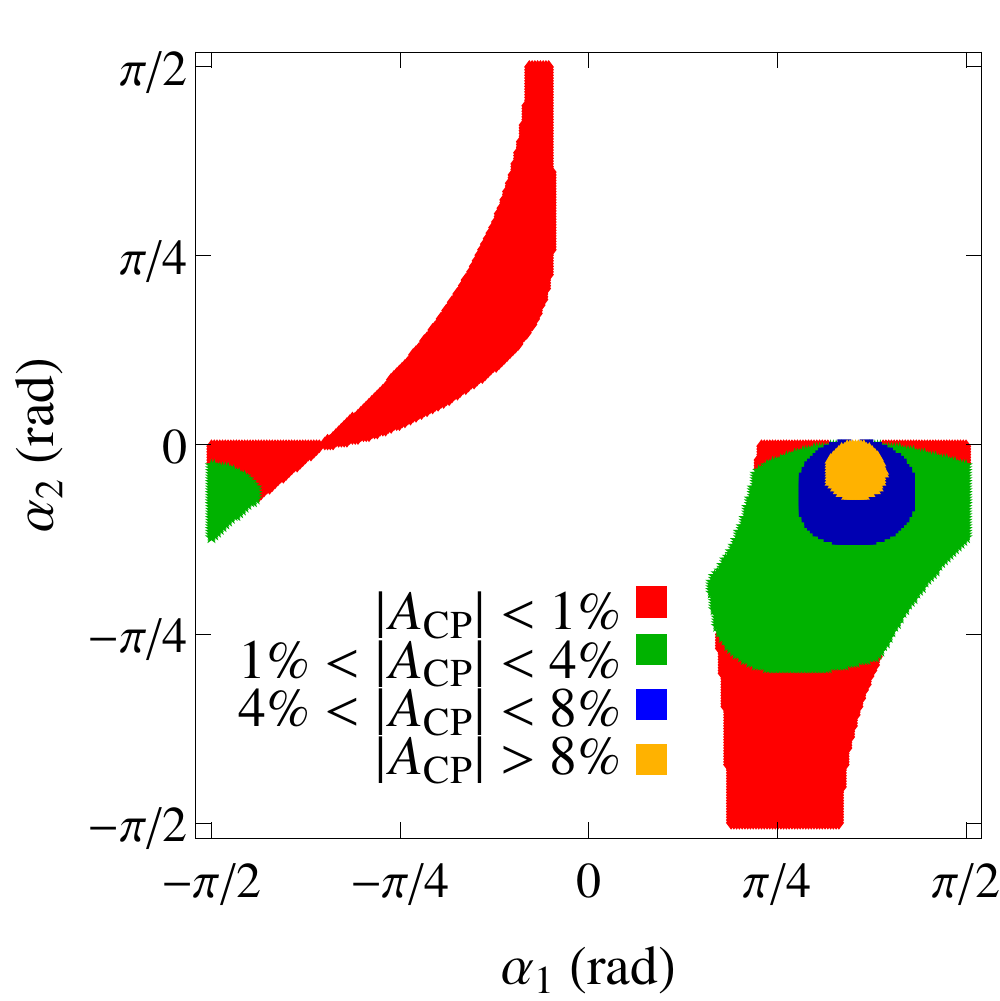} \\
\includegraphics[height=3.0in,width=7cm]{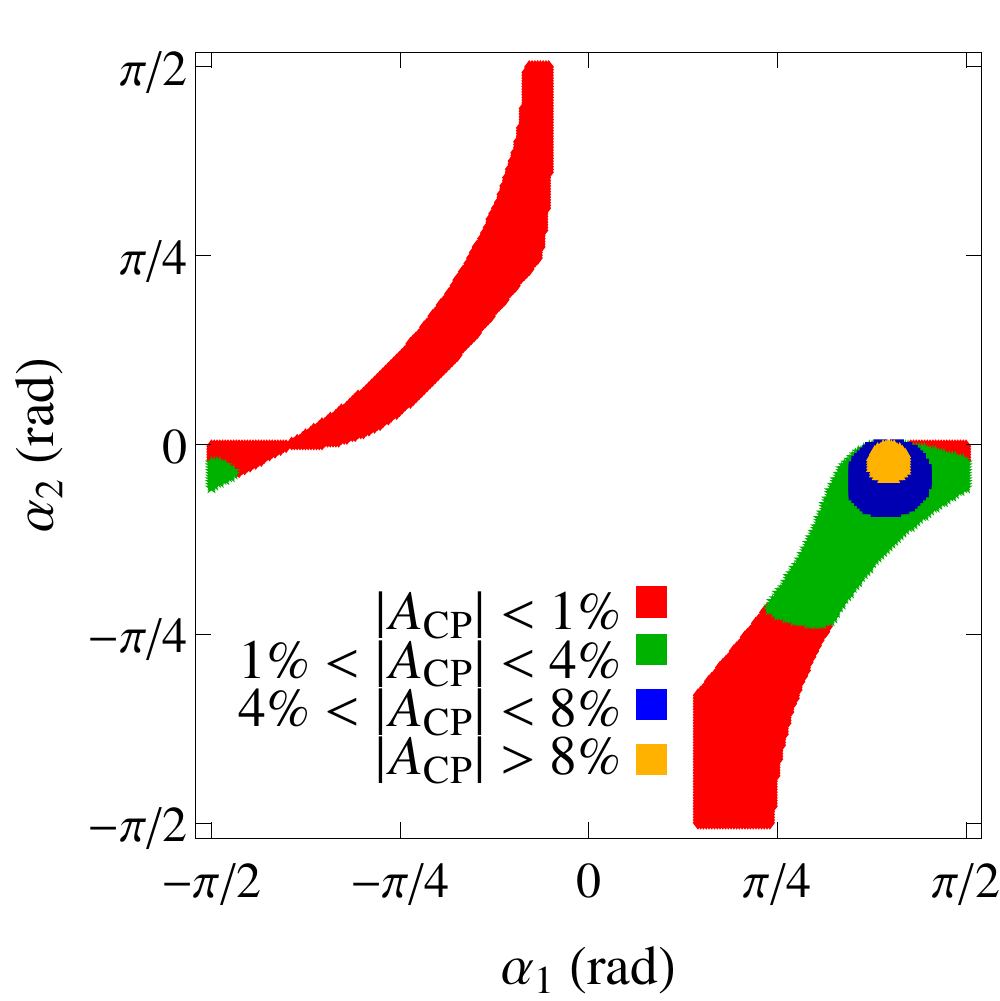}
\hskip 0.1cm
\includegraphics[height=3.0in,width=7cm]{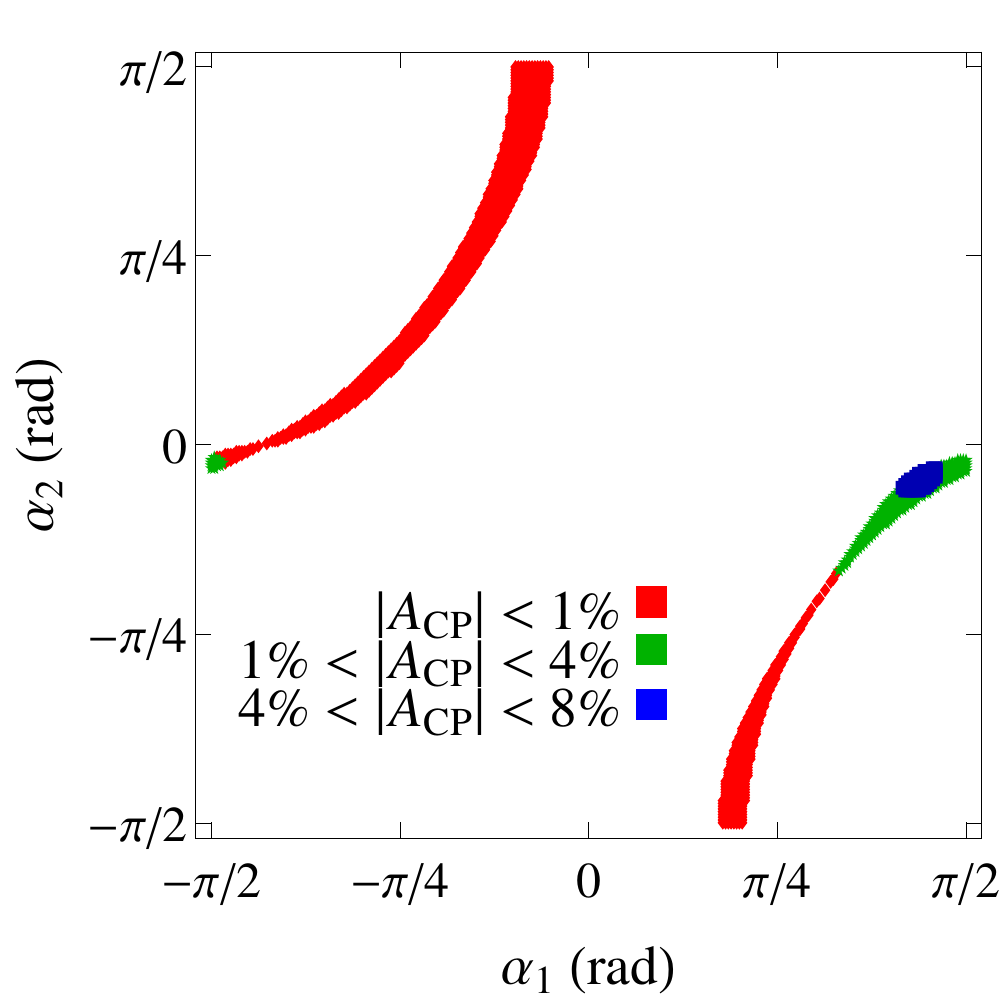}
\caption{The allowed parameter regions in the $(\alpha_1,\alpha_2)$ plan
in the C2HDM together with the absolute value of the CPV asymmetry $A_{D,WH_1^0}^{CP}$.
The other parameters are the same
as for figure \ref{scan1}.}
\label{scan2}
\end{figure}
\begin{figure}[!ht]
\centering
\includegraphics[height=3.0in,width=7cm]{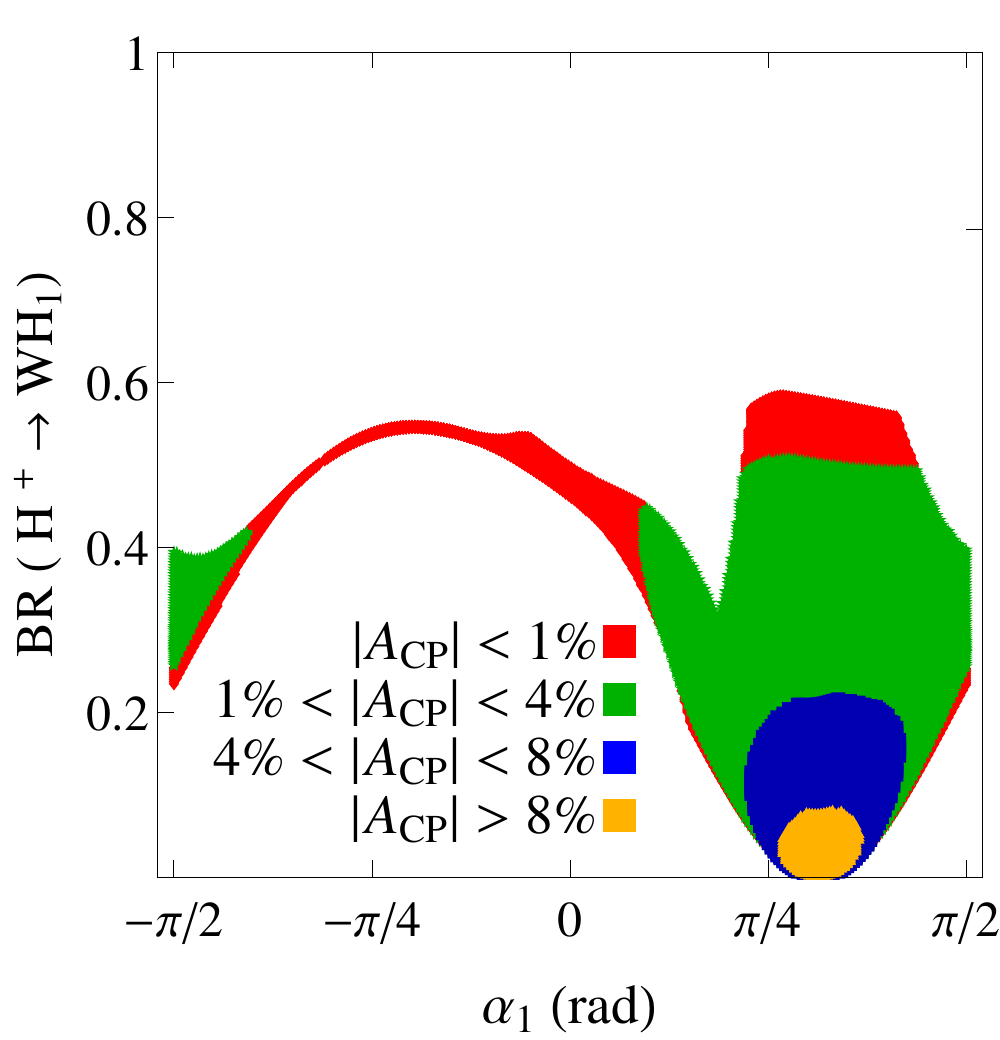}
\hskip 0.1cm
\includegraphics[height=3.0in,width=7cm]{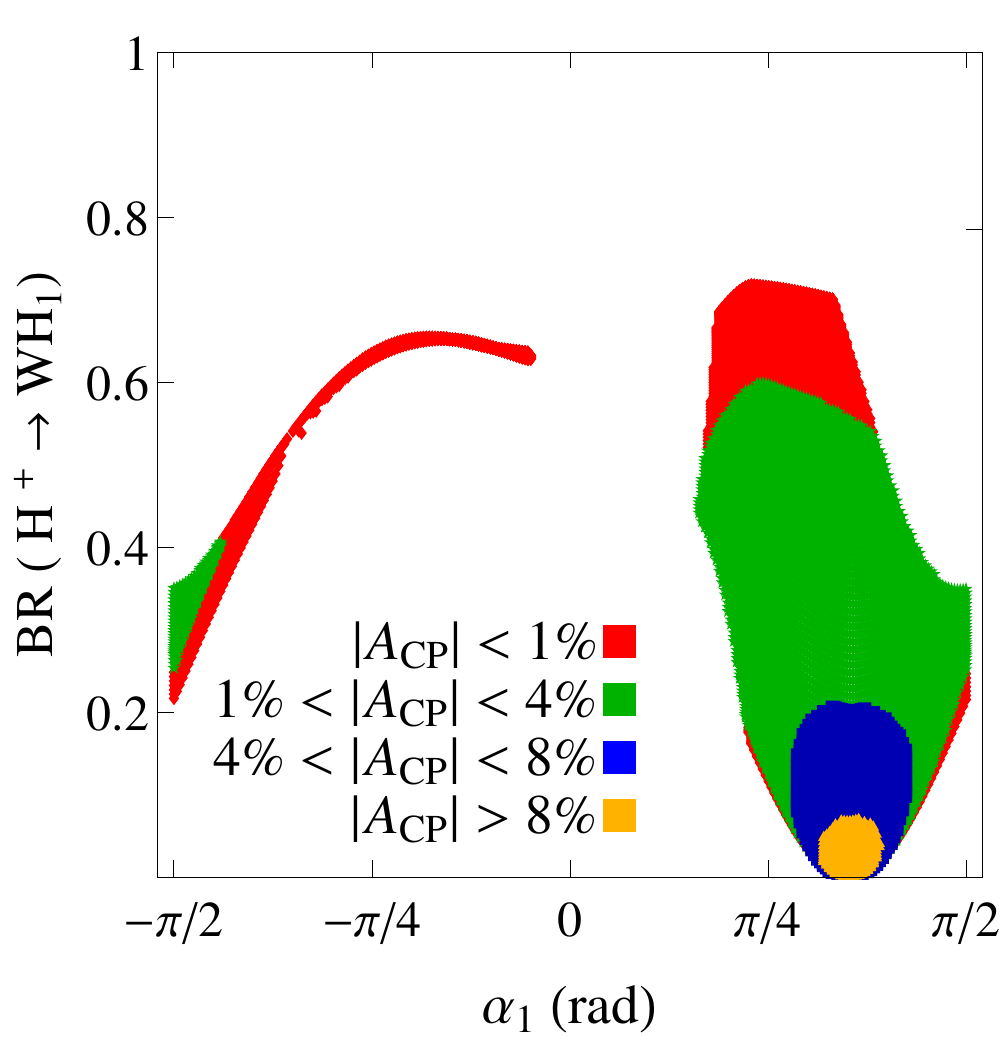} \\
\includegraphics[height=3.0in,width=7cm]{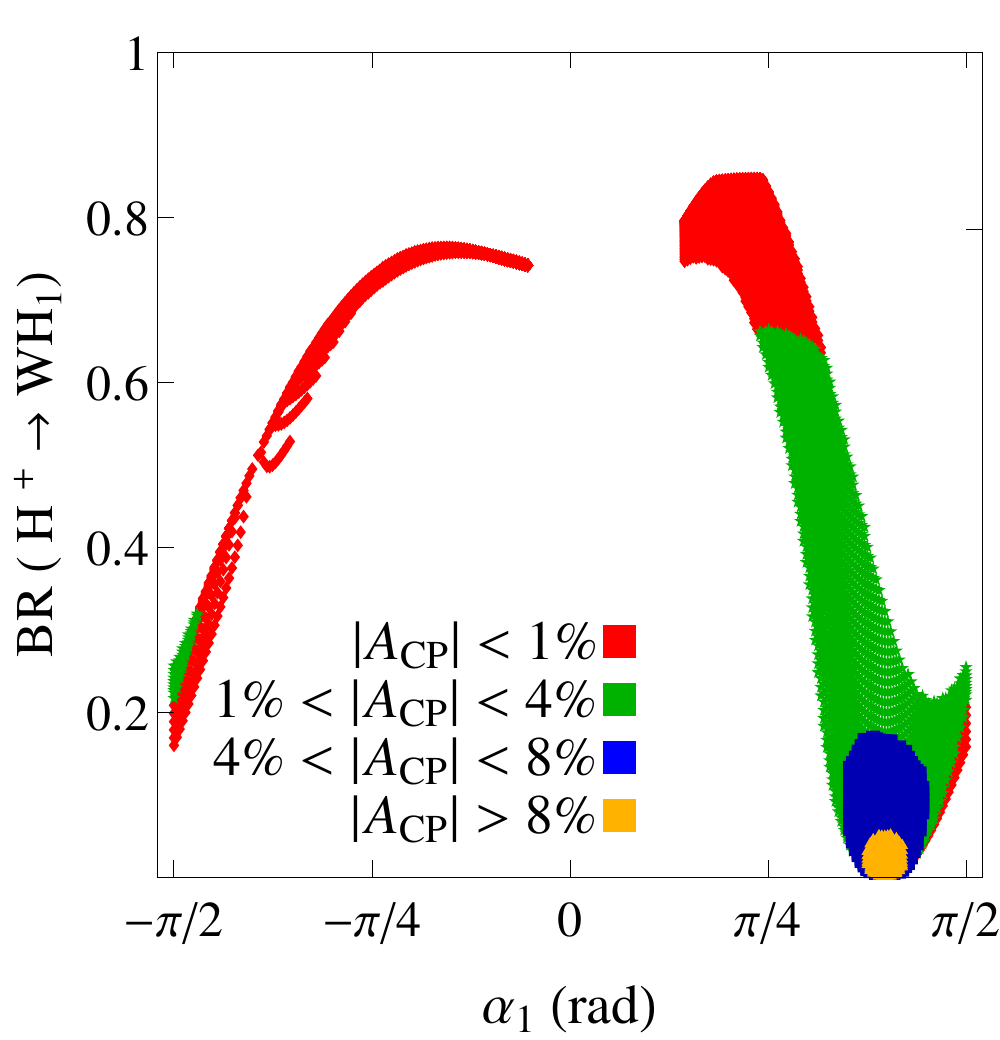}
\hskip 0.1cm
\includegraphics[height=3.0in,width=7cm]{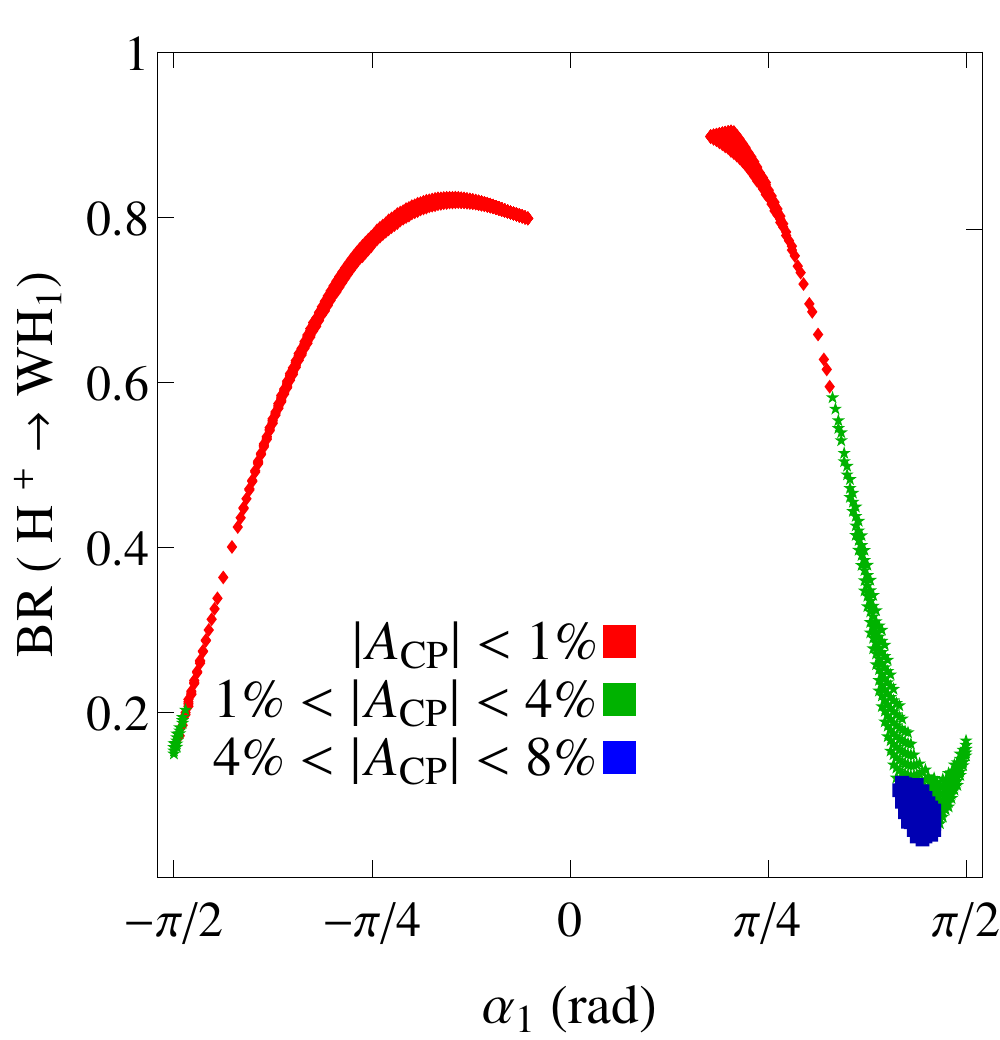}
\caption{The BR$(H^+ \to W^+ H_1^0)$ as a function of $\alpha_1$ with $\alpha_2$ in
the allowed parameter range. The other parameters are the same
as for figure \ref{scan1}.}
\label{scan3}
\end{figure}

In figure \ref{scan2} we show a similar scan for the $H^\pm \to W^\pm H_1^0$ decay.
When we scan over $\alpha_1$ and $\alpha_2$
for a fixed value of $\alpha_3$, we find a region where the coupling ${\cal C}(H_1^0 H^\pm W^\mp)$
becomes very small. It is seen in the figure that
for $\tan\beta=1.5$
${\cal C}(H_1^0 H^\pm W^\mp) \to 0$ for $\alpha_2\approx 0$ and
$\alpha_1\approx \pi/3$ (yellow color).
The position of this region shifts to the right for larger values of $\tan\beta$.
The coupling ${\cal C}(H_1^0 H^\pm W^\mp )$ is complex. Therefore,
the condition ${\cal C}(H_1^0 H^\pm W^\mp ) \to 0$ also implies that its imaginary part
$R_{13}\to 0$, which furthermore means that $H_1^0$ is dominated by its CP-even component.

This can be seen also from the sum rule given by eq. (\ref{sum1}),
which tell us that if \\
$|{\cal C}(H_1^0 H^+ W^- )|^2 \to 0$  then
${\cal C}(H_1^0 WW)^2 \to 1$ and $H_1^0$, which has a maximal coupling to
a W pair, is dominated by its CP-even component.
According to our definitions of the CPV asymmetries in section \ref{sec:CPasym}, when
$|{\cal C}(H_1^0 H^\pm W^\mp)|^2 \to 0$,
the tree-level width $\Gamma^{\rm tree}(H^\pm \to W^\pm H_1^0 )\to 0$ and
consequently the CPV asymmetry will increase considerably. However, the large rate asymmetry
we would obtain in this case, would go together with a small BR
of the $H^\pm \to W^\pm H_1^0$ decay.

In figure \ref{scan3} we show the BR of
$H^+  \to W^+ H_1^0$ as a function of $\alpha_1$
for various values of $\tan\beta$ and the other parameters
fixed as in figure \ref{scan1}. It can be seen in the plots that in some cases
the BR of  $H^+ \to W^+ H_1^0$ can be larger than $80\%$
together with a CPV asymmetry of a few percent.

\subsection{CP violation in $H^\pm$-production}

\begin{figure}[!ht]
\begin{tabular}{cc}\hspace*{-1cm}
\resizebox{85mm}{!}{\includegraphics{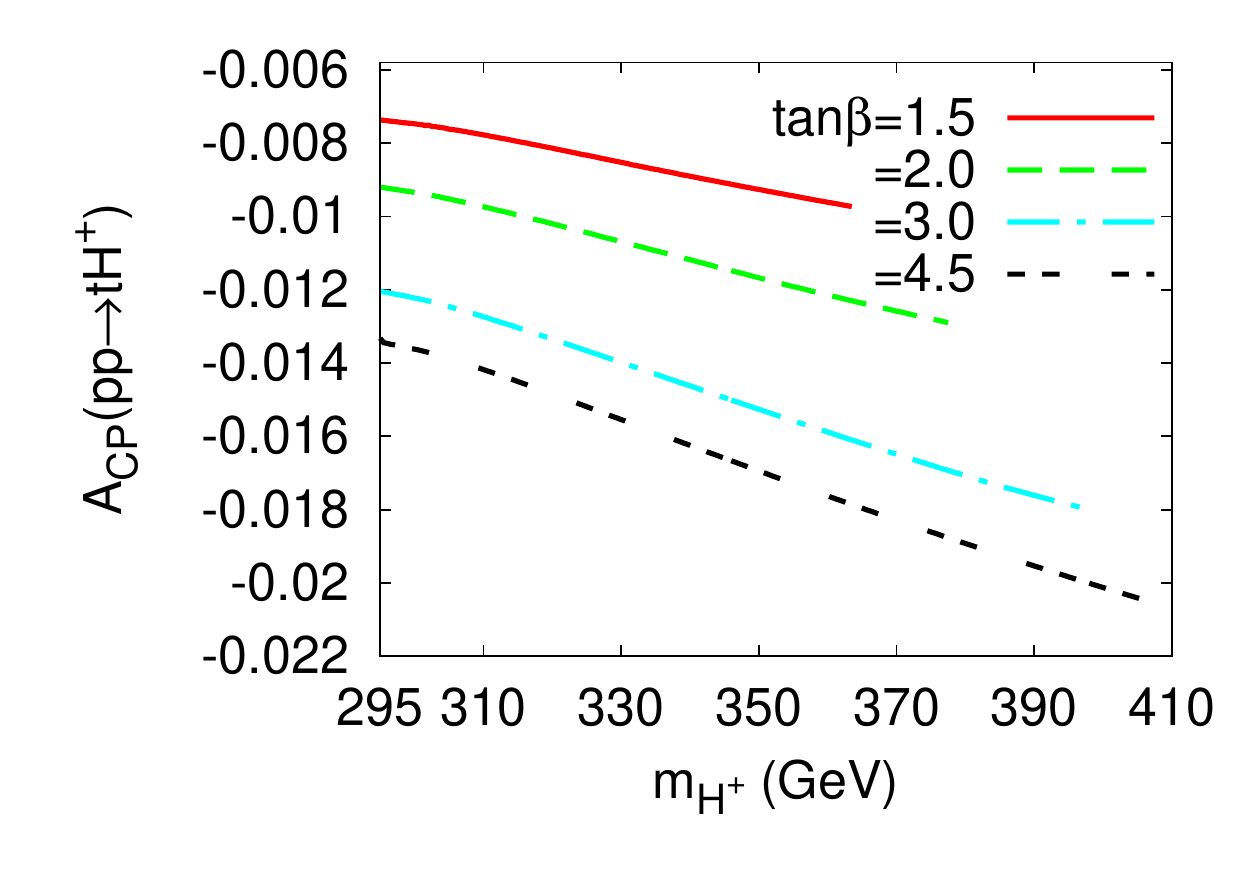}} & \hspace{-1.cm}
\resizebox{85mm}{!}{\includegraphics{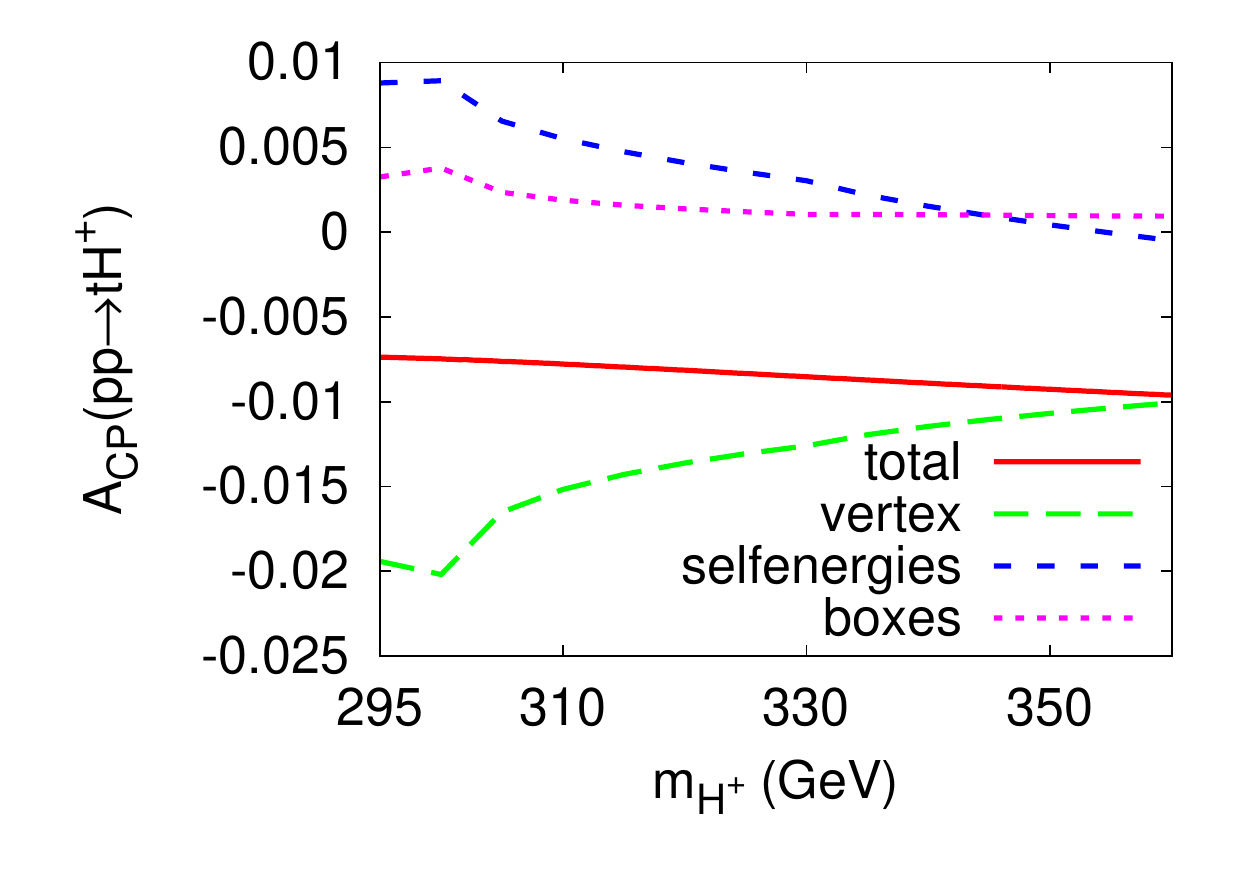}}
\end{tabular}
\caption{CPV asymmetry $A_P^{CP}$  (\ref{CPVprod}); Left: as a
function of charged Higgs mass for four values of $\tan\beta$ and the other
parameters are as in figure \ref{Hptotb}: $M_{H_1^0}= 120$ GeV, $M_{H_2^0}= 220$ GeV,
${\rm Re} (m_{12})=170$ GeV, $\alpha_1=0.8$, $\alpha_2=-0.9$ and $\alpha_3=\pi/3$;
Right: vertex, selfenergy and box contributions to $A_P^{CP}$ as a function of $M_{H^+}$ for $\tan \beta = 1.5$.} \label{prodasymm}
\end{figure}

We also study the CPV asymmetry $A^{CP}_P$, given by
eq. (\ref{CPVprod}), for charged Higgs production
 (\ref{prod2}) at the LHC with $\sqrt{s}=14$~TeV. For a direct comparison with the results of section \ref{sec:numdecays},
in figure \ref{prodasymm} we
present the asymmetry $A^{CP}_P$ as a function of the charged
Higgs mass $M_{H^+}$ for the same parameter
set as used for figure \ref{Hptotb}. One can see that here the asymmetry
is of the same order of magnitude or smaller than in the $H^\pm \to
t b$. The absolute value increases with $\tan \beta$, and can go up to 2\% for $\tan \beta=4$ and
$M_{H^+}=400$ GeV. But in contrast to the $H^\pm \to t b$ decay the production asymmetry is negative.
Therefore, the total asymmetry $A^{CP}_P + A^{CP}_{D, tb}$ is very small.
However, $A^{CP}_P$ has the same sign like
$A^{CP}_{D,WH_{1,2}^0}$ and according to eq. (\ref{finalf}) this will increase
the total asymmetry in these bosonic modes.

In figure {\ref{prodasymm}} we show the vertex, selfenergy and
box contributions to $A^{CP}_P$ for $\tan \beta=1.5$. The vertex contribution is negative and dominating, especially for
smaller masses of the charged Higgs. The selfenergies and box
contributions are both positive, but their sum partially cancels
with the contributions of the vertex diagrams.
\begin{figure}[!ht]
\centering
\includegraphics[height=3.0in,width=7cm]{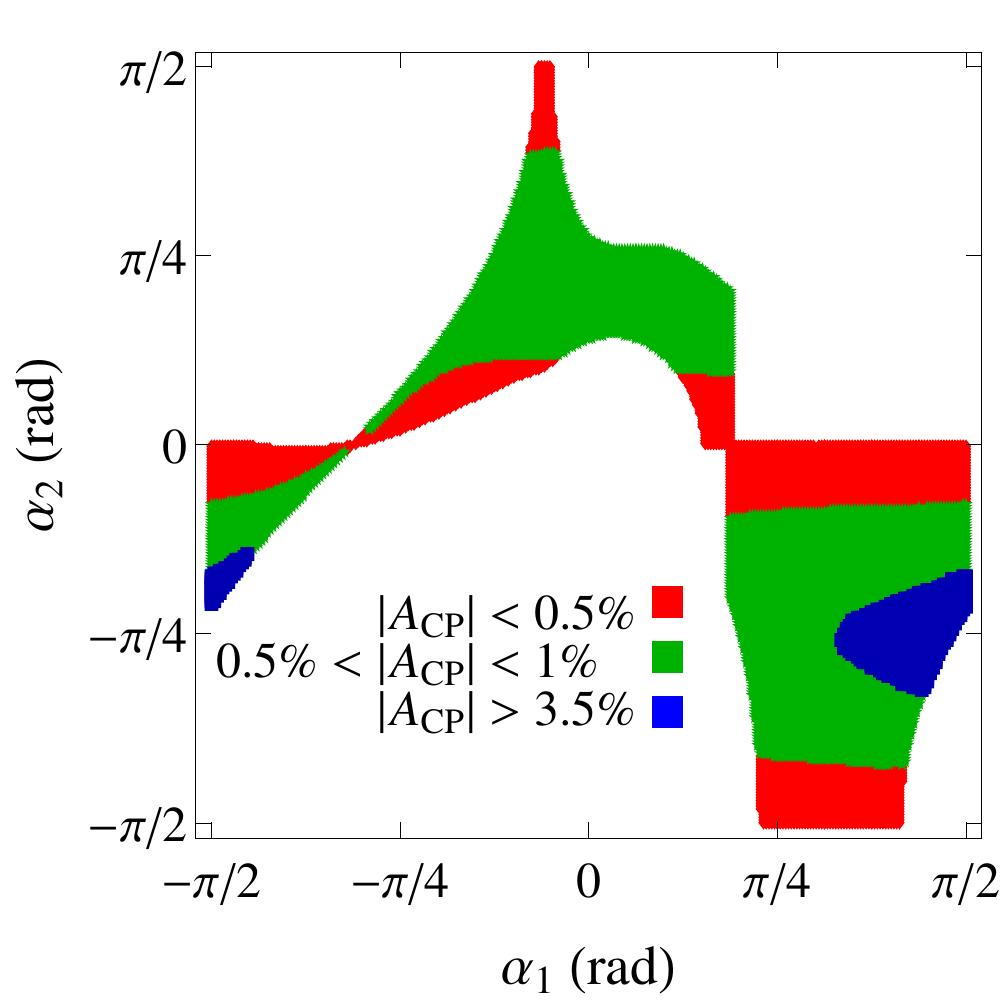}
\hskip 0.1cm
\includegraphics[height=3.0in,width=7cm]{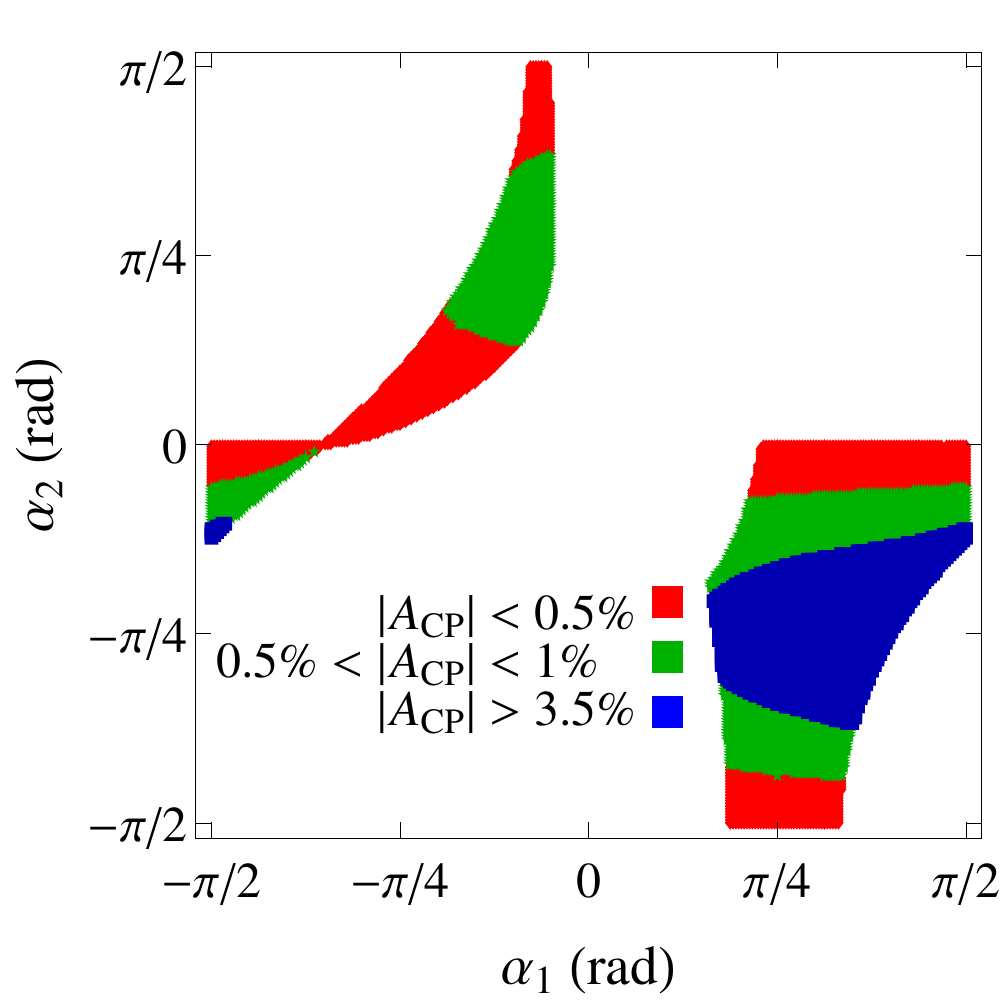} \\
\includegraphics[height=3.0in,width=7cm]{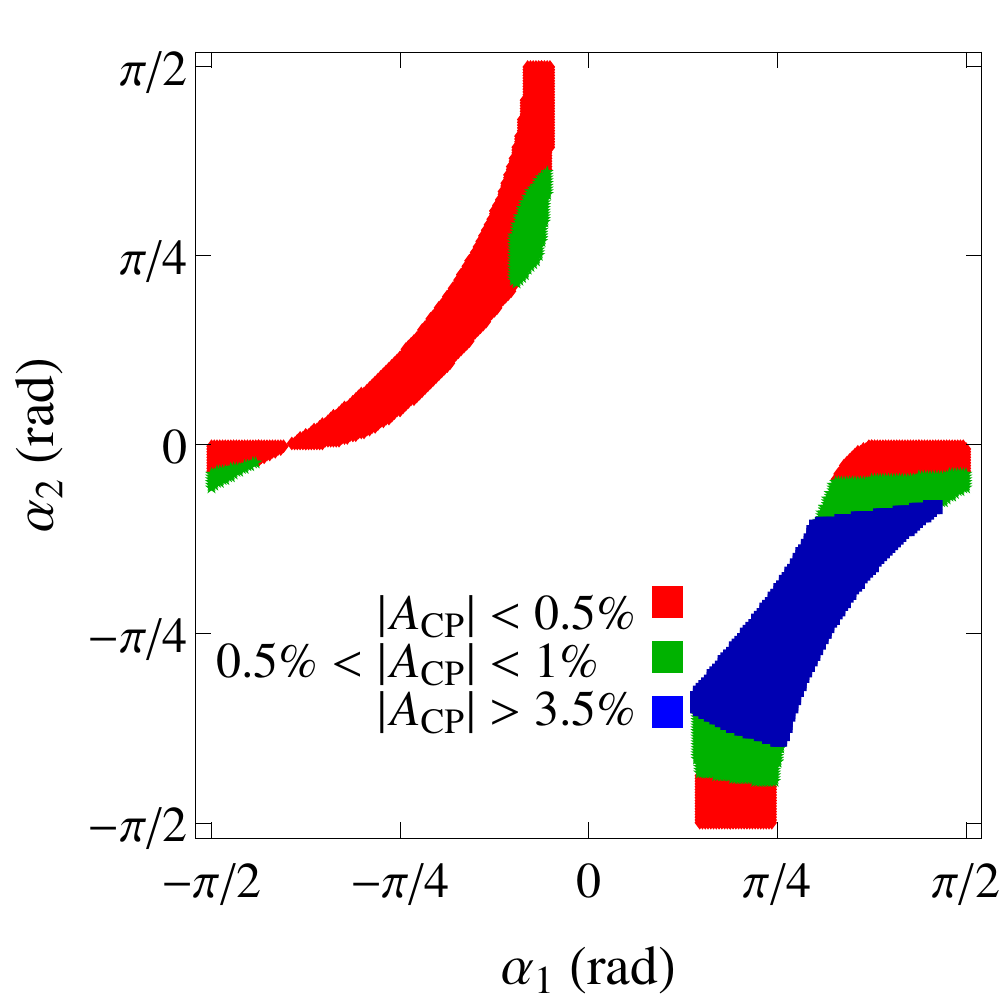}
\hskip 0.1cm
\includegraphics[height=3.0in,width=7cm]{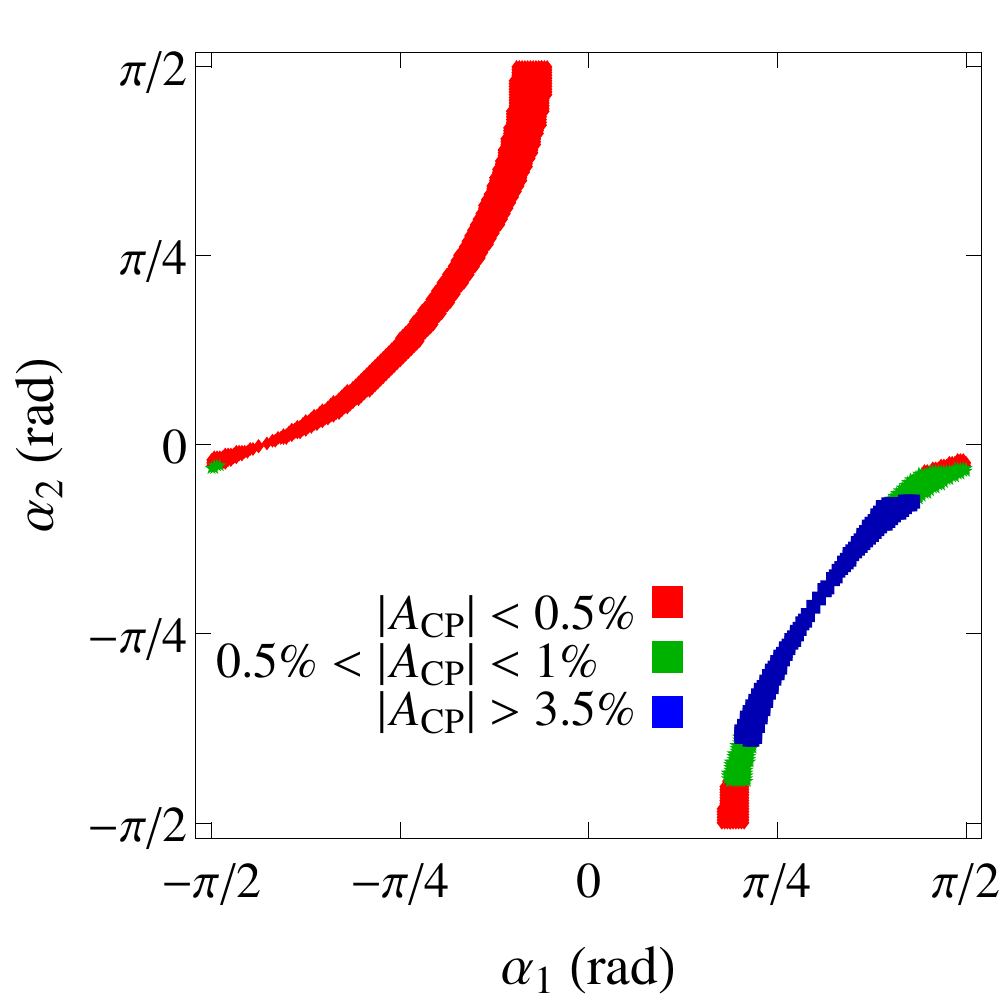}
\caption{The allowed parameter regions in the $(\alpha_1,\alpha_2)$ plan
in the C2HDM together with the absolute value of the CPV asymmetry $A_{P}^{CP}$.
The other parameters are the same as in figure \ref{scan1}.}
\label{scan4}
\end{figure}

In figure \ref{scan4} we show a scan over the angles $\alpha_1$ and $\alpha_2$, together with the absolute value
of the asymmetry $A_P^{CP}$, as for the decays in the previous section, see figure \ref{scan1} and figure \ref{scan2}.
The plots are very similar in size and the asymmetry can reach $\sim$ 2\%.

In order to have a better overview on the CPV asymmetry in a larger
parameter space and having in mind that the integration over the
PDFs slows down the calculation considerably, we perform a numerical scan
over the C2HDM parameters  (\ref{paramm}) using Grid computing:
\begin{eqnarray}
M_{H_1^0} & = &115 \div 125~{\rm GeV},~{\rm with~step~size}~5~{\rm
GeV}\,,\nonumber
\\
 M_{H_2^0} & = & 150 \div 400~{\rm GeV},~{\rm with~step~size}~50~{\rm
GeV}\,, \nonumber
\\
 M_{H^+}  & = & 300 \div 550~{\rm GeV},~{\rm with~step~size}~25~{\rm GeV}\,,
\nonumber
\\
 {\rm Re}\, (m_{12}) & = & 10 \div 460,~{\rm with~step~size}~50\,,
\nonumber
\\
 \tan \beta & = & 1 \div 8,~{\rm with~step~size}~1\,, \nonumber
\\
 \alpha_1 & = & \pi/2 \div \pi/2,~{\rm with~step~size}~\pi/9\,,\nonumber
\\
 \alpha_2 & = & \pi/2 \div \pi/2,~{\rm with~step~size}~\pi/9\,, \nonumber
\\
 \alpha_3 & = & 0 \div \pi/2,~{\rm with~step~size}~\pi/9\,, \nonumber
\\
 \label{scan}
\end{eqnarray}
\begin{figure}[h!]
\begin{center}
\begin{tabular}{cc}
\resizebox{7.0cm}{!}{\includegraphics{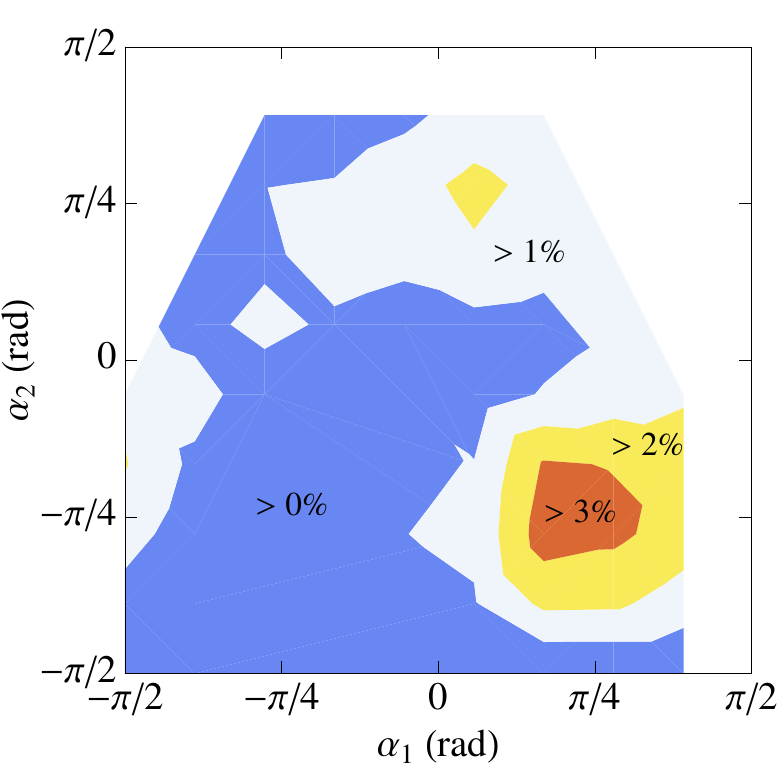}} &
\hspace{1.cm}
\resizebox{6.5cm}{!}{\includegraphics{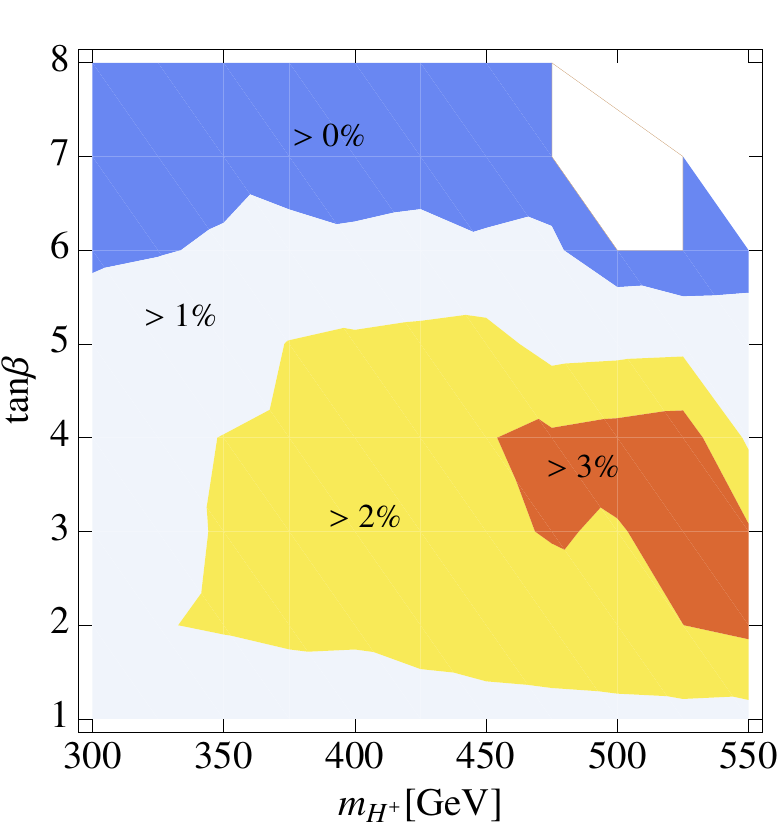}}
\end{tabular}
\end{center}
\caption{The maximal value of the asymmetry $|A_P^{CP}|$ as a function of $\a_1$ and $\a_2$ -- left; and $M_{H^+}$
and $\tan \b$ -- right, based on the parameter scan  (\ref{scan}).}
\label{numscan2}
\end{figure}

The total number of the scanned parameter points is quite
large: 6415200. After having applied the theoretical and experimental constraints, see
section \ref{sec:th.constr.} and section \ref{sec:exp.constr.}, the allowed number of parameter points relegates to
67861, which is only $1 \%$ of the scanned parameter space.

The numerical results based on the parameter scan  (\ref{scan}) show that in most of the
cases the asymmetry is very small, practically zero. The non-zero
asymmetry is distributed in two non-symmetric bunches - to the right (up to $\sim$ 2\%) and to the left (down to $\sim -3$\%) from the zero. The non-symmetric
structure of the distribution is due to the asymmetric $\Delta \rho$ constraints, see section \ref{sec:exp.constr.}.
Furthermore, the results from the numerical scan  (\ref{scan}) show that the CPV asymmetry
$A^{CP}_P$ can reach 3.5 \% (negative) for single isolated points of
the parameter space.

In figure \ref{numscan2} we show the maximal value of $|A_P^{CP}|$ as a function of $\a_1$ and $\a_2$ -- left; and $M_{H^+}$
and $\tan \b$ -- right, based on the parameter scan  (\ref{scan}). One sees small regions where the asymmetry
can go up to
$\sim$ 3\%. These are roughly: $(\a_1, -\a_2)\in (\pi/6\div \pi/3)$ and relatively large
$M_{H^+}$ and $\tan \beta$:
$M_{H^+} \in ({460 \div 550})$ and $\tan \beta  \in ({2 \div 4})$. However, our experience showed that making a plot
in such region is a question of fine-tuning of the parameters due to the severe theoretical constraints ( see section \ref{sec:th.constr.} ), which  often cut the parameter space into unconnected subspaces.

\subsection{CP violation in $H^\pm$-production and subsequent decays}

Combining the production process  (\ref{procon}) with the subsequent decay $H^\pm \to tb$, in \cite{CPVinH+tMSSM} we have shown analytically that
the charged Higgs selfenergy contributions to the total asymmetry
from the production and the decay exactly
cancel. Moreover, we have shown that in the MSSM the vertex contributions from the production and from the decay
also partially cancel. Eventually, the main
contribution in the total asymmetry comes from the MSSM box contributions to the production.
\begin{figure}[!ht]
\begin{center}
\begin{tabular}{cc}\hspace*{-1cm}
\resizebox{85mm}{!}{\includegraphics{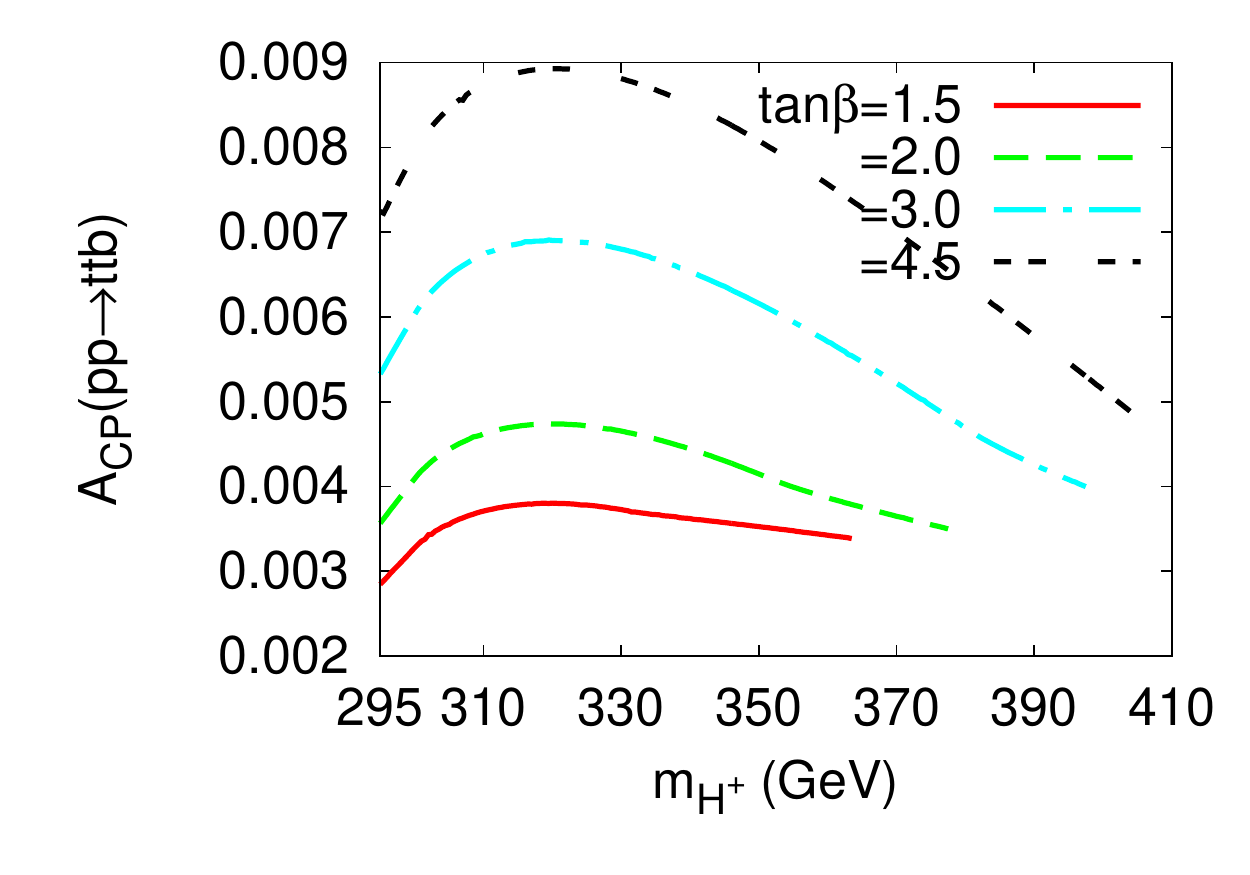}}
\end{tabular}
\end{center}
\caption{CPV asymmetry $A_{t\bar{b}}^{CP}$  (\ref{AfCP})
as a function of the charged Higgs mass for four values of $\tan\beta$
and the other parameters are as in figure \ref{Hptotb}: $M_{H_1^0}= 120$ GeV, $M_{H_2^0}= 220$ GeV,
${\rm Re} (m_{12})=170$ GeV, $\alpha_1=0.8$, $\alpha_2=-0.9$ and $\alpha_3=\pi/3$.}
\label{prod+decaytb}
\end{figure}
\begin{figure}[!ht]
\begin{tabular}{cc}\hspace*{-1cm}
\resizebox{85mm}{!}{\includegraphics{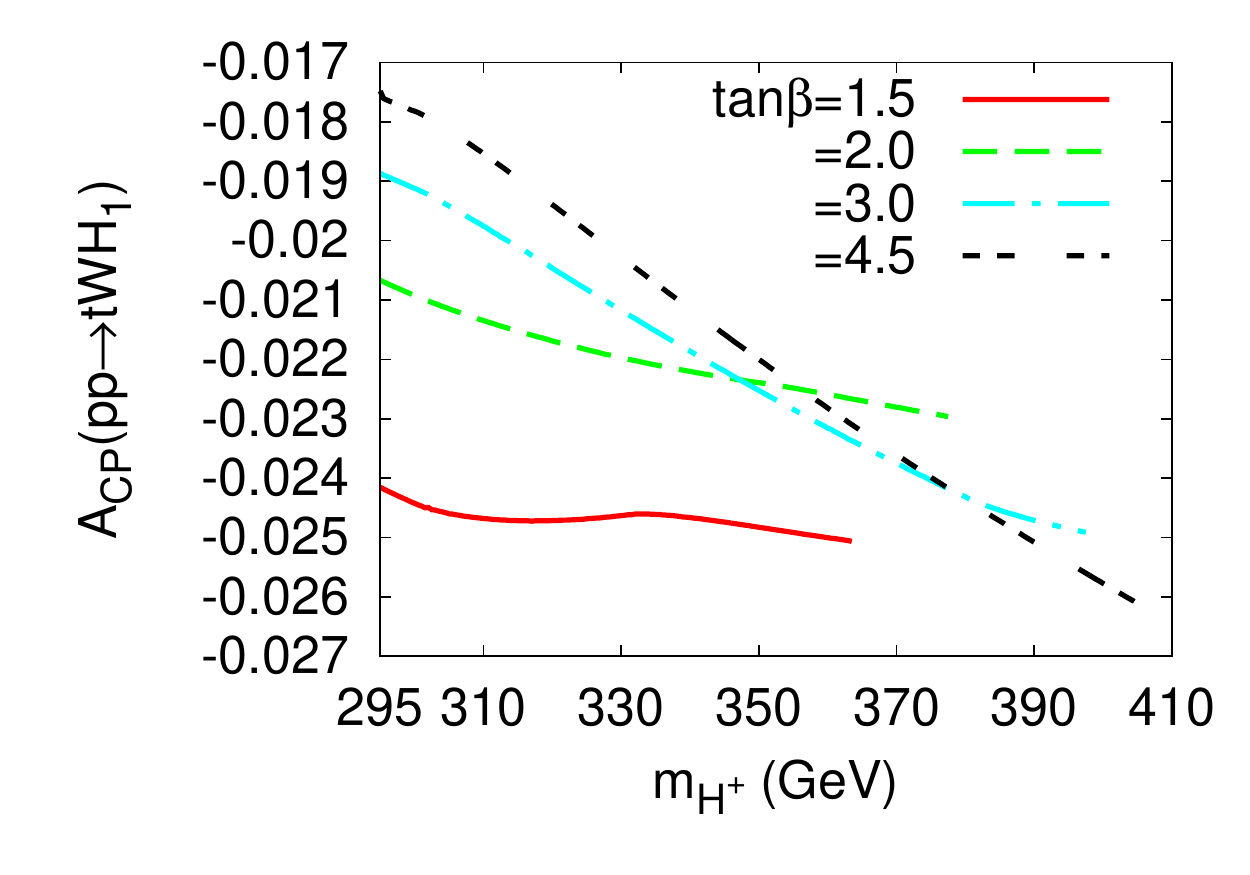}} & \hspace{-1.cm}
\resizebox{85mm}{!}{\includegraphics{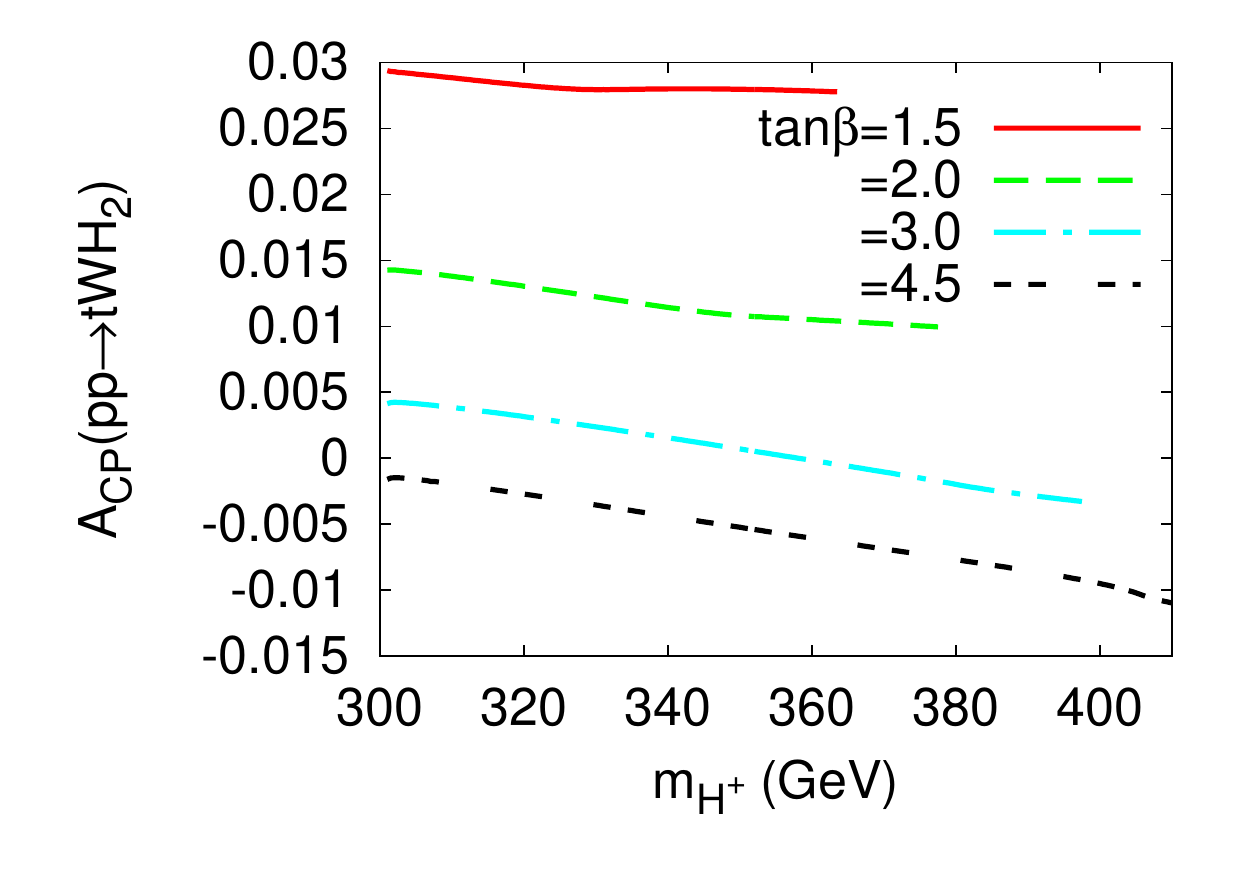}}
\end{tabular}
\caption{CPV asymmetries $A_{W H_1}^{CP}$ and $A_{W H_2}^{CP}$  (\ref{AfCP})
as functions of $M_{H^+}$ for four values of $\tan\beta$.
The other parameters are: $m_{H_1^0}= 120$ GeV, $m_{H_2^0}= 220$ GeV,
$m_{12}=170$ GeV, $\alpha_1=0.8$, $\alpha_2=-0.9$,  and $\alpha_3=\pi/3$.}
\label{prod+decaysWH12}
\end{figure}

In this paper we study the same production process, but in the C2HDM. The selfenergy cancellations observed in \cite{CPVinH+tMSSM}
are not model dependent and we expect them to occur again.
In figure \ref{prod+decaytb} we show the asymmetry $A_{tb}^{CP}$ as a function of the charged Higgs mass.
One sees that the total asymmetry is of an order of magnitude smaller in comparison to the individual decay and production asymmetries.
The cancellations are easily traced back comparing the selfenergy and vertex contributions in figure \ref{Hptotb}
and figure \ref{prodasymm}, which are of the same magnitude, but with opposite signs.
However, no such
cancellations occur in the bosonic modes: $H^\pm \to W^\pm
H_{1,2}^0$. In figure \ref{prod+decaysWH12} we show the asymmetries $A_{W H_1}^{CP}$ and $A_{W H_2}^{CP}$  (\ref{AfCP})
as functions of $M_{H^+}$ for four values of $\tan\beta$. As one can see, here the effect is bigger and the asymmetry $A_{W H_1}^{CP}$
can reach up to 2.5\% (negative), while the asymmetry $A_{W H_2}^{CP}$ can go up to 3\% (positive).

In the analogous study performed in the MSSM \cite{CPVinH+tMSSM} the main contribution for the case with the decay $H^\pm \to t b$ is due to box graphs with gluino exchange
which involves the strong coupling constant. This asymmetry reaches its largest value of $\sim$ -12\% for $m_{H^+} \sim 550$~GeV with a branching ratio BR($H^+ \to t \bar b$) $\sim$ 20\%.

In contrast to  \cite{CPVinH+tMSSM},
in the present study the asymmetries are due to the exchange of neutral Higgs bosons and therefore of electroweak nature.
In the $t b$ mode the asymmetry always remains below $\sim$ 1\% with BR($H^+ \to t \bar b$) $\sim$ 50\% for $m_{H^+} \sim 350$~GeV.
In the $W^\pm H_1$ mode it can reach 2.5\% for $m_{H^+} \sim 350$~GeV with BR($H^+ \to W^\pm H_1$) up to $\sim$ 45\%.\\

The production rate of $H^+$ at the LHC for $m_{H^+} = $~350~GeV
and $\tan\beta = 1.5$ is $\sim 450$~fb (including a K-factor from QCD
of 1.3, see \cite{Belyaev:2002eq}).
Assuming a branching ratio of 50\% and an integrated luminosity of 200~fb$^{-1}$ we get
$N \sim 45000$ and $\sqrt{N} \simeq 212$. Furthermore, assuming
$A^{CP} \sim 0.02$, the statistical significance
$\sqrt{N} A^{CP}_{t b} \sim 4.2$. However, in a realistic study the
actual signal production rate will be most likely reduced.
Moreover, the total background rate for charged Higgs
production at the LHC is quite large \cite{Belyaev:2002eq}. The obtained statistical significance
might be too low for a clear observation in the first stage of the LHC and therefore
an upgrade of its luminosity would be necessary to await. 

\section{Conclusions}

In the type II complex 2HDM with softly broken $Z_2$ symmetry, the non-zero and complex $m_{12}^2$ parameter of the tree-level Higgs potential gives rise to CP violation in the production process $pp \to H^\pm t+X$,
and in the dominant decay modes
of $H^\pm$ to $tb$, and to $W H_i$, i=1,2. We have calculated
the corresponding CP violating rate asymmetries at one-loop level both in the production and in the
decays, as well as in the combined processes.
A detailed numerical analysis has been performed. The dependences of the asymmetries on
the C2HDM parameters are studied taking into account the theoretical constraints, experimental lower
bound on the charged Higgs mass from $B \to X_s \gamma$ and the constraint on the $\rho$ parameter.

The calculations have been performed with the help of the packages FeynArts and FormCalc.
For that purpose,
a new model file for FeynArts has been created and
corresponding fortran drivers for FormCalc have been written.

In the allowed parameter space
of the C2HDM parameters, which is severely constrained by vacuum stability, perturbativity, unitarity,
lower charged Higgs mass bound and $\Delta \rho$,
the studied CP violating asymmetries cannot be greater than $\sim$3 \%.
This is in contrast with a similar
study performed within the MSSM \cite{CPVinH+tMSSM}. There the CPV asymmetries can be larger by an order of magnitude.
However, after having taken into account the relevant branching ratios and cross sections, the measurability of the
studied CPV asymmetries in the C2HDM and in the MSSM
at the LHC has roughly the same statistical significance,
which is maybe not big enough for a clear observation at the LHC. At the
SLHC with a design luminosity bigger by a factor of $\sim$~10, such
a measurement would be worth of being performed.


\appendix
\section{The $\Delta \rho$  constraint}
\label{App:rho}
The $\rho$ parameter is defined by the ratio of the neutral and charged
currents at vanishing momentum transfer. In the C2HDM, at tree level
$\rho=m_W^2/(c_W^2m_Z^2)=1$ which is in perfect agreement with the
experimentally measured value. This relation may be spoiled by
radiative corrections if they are large.

For the extra contributions $\Delta\rho$ to the $\rho$ parameter from the additional
scalars Higgses in the C2HDM, we have implemented in our code the
expression given in \cite{ElKaffas} in terms of physical masses
and elements $R_{jk}$ of the rotation matrix  (\ref{matRo}).
The analytic expression for $\Delta \rho$ can be split into two contributions:\\
i) Higgs--Higgs contribution (HH):\footnote{The relevant couplings are given explicitly in
appendix~B of \cite{ElKaffas}.}
\begin{eqnarray}\hspace*{-0.8cm}
A_{WW}^{HH}(0)-\cos^2\theta_W\, A_{ZZ}^{HH}(0)
={g^2\over 64\pi^2}\sum_j\Bigl[ \{[\sin\beta R_{j1}-\cos\beta R_{j2}]^2 +R^2_{j3}\} F_{\Delta\rho}(M_{H^\pm}^2,M_j^2)&& \nonumber\\
-\sum_{k>j}[(\sin\beta R_{j1}-\cos\beta R_{j2})R_{k3}
-(\sin\beta R_{k1}-\cos\beta R_{k2})R_{j3}]^2\,F_{\Delta\rho}(M_j^2,M_k^2) \Bigr]\,,\quad &&
 \label{Eq:deltarho-HH}
\end{eqnarray}
with
\be
F_{\Delta\rho}(m_1^2,m_2^2)=\half(m_1^2+m_2^2)
-\frac{m_1^2m_2^2}{m_1^2-m_2^2}\log\frac{m_1^2}{m_2^2}.
\ee
ii) Higgs--ghost contribution (HG):
\begin{eqnarray}
A_{WW}^{HG}(0)-\cos^2\theta_W\, A_{ZZ}^{HG}(0)
=\frac{g^2}{64\pi^2}\Bigl[\sum_j[\cos\beta\, R_{j1}+\sin\beta\, R_{j2}]^2 \times \qquad &&\nonumber\\
\times \Bigl(3F_{\Delta\rho}(M_Z^2,M_j^2)-3F_{\Delta\rho}(M_W^2,M_j^2)\Bigl)
\quad +3F_{\Delta\rho}(M_W^2,M_0^2)-3F_{\Delta\rho}(M_Z^2,M_0^2)\Bigr]\,.
\label{Eq:deltarho-HG}
\end{eqnarray}
From this contribution one
has to substract the SM Higgs contribution of a mass $M_0$ \cite{ElKaffas}.
The choice of $M_0$ is usually consistent with  the fit analysis. As a default value we take $M_0=120$ GeV.

\section{Expressions for $\l_i$ }
\label{App:lambda}

In order to work with the parameter set  (\ref{paramm}), we need the explicit relations between the
parameters of the Higgs potential and the physical Higgs masses and rotation angles.
It is straightforward to derive the
 $\lambda_i$, $i=1,2,3,4,5$ as functions of our
input parameters  (\ref{paramm}) \cite{Per}:
\begin{eqnarray}
\lambda_1 & = & \frac{1}{c_\beta^2 v^2}[c_1^2 c_2^2 M_{H_1^0}^2+(c_1 s_2
s_3+s_1 c_3)^2 M_{H_2^0}^2+(c_1 s_2 c_3 -s_1 s_3)^2 M_{H_3^0}^2-s_\beta^2
\mu^2]\,, \nonumber \\
\lambda_2& = &\frac{1}{s_\beta^2 v^2}[s_1^2
c_2^2 M_{H_1^0}^2+(c_1 c_3- s_1 s_2 s_3)^2 M_{H_2^0}^2+(c_1 s_3+ s_1 s_2
c_3)^2 M_{H_3^0}^2-c_\beta^2 \mu^2]\,,\nonumber \\
\lambda_3& = &\frac{1}{s_\beta c_\beta v^2}\{c_1 s_1[c_2^2
M_{H_1^0}^2+(s_2^2 s_3^2- c_3^2) M_{H_2^0}^2+( s_2^2 c_3^2 - s_3^2) M_{H_3^0}^2]
\nonumber \\ &&+s_2 c_3
s_3(c_1^2-s_1^2)(M_{H_3^0}^2-M_{H_2^0}^2)\}+\frac{1}{v^2}[2M_{H^+}^2-\mu^2]\,,
\nonumber \\
\lambda_4& = &\frac{1}{v^2}[s_2^2 M_{H_1^0}^2+c_2^2 s_3^2
M_{H_2^0}^2+c_2^2 c_3^2 M_{H_3^0}^2+\mu^2- 2M_{H^+}^2]\,, \nonumber \\
{\rm Re}(\lambda_5)& = &\frac{1}{v^2}[-s_2^2 M_{H_1^0}^2-c_2^2 s_3^2 M_{H_2^0}^2-c_2^2
c_3^2 M_{H_3^0}^2+\mu^2]\,, \nonumber \\
{\rm Im}(\lambda_5)& = &-\frac{1}{c_\beta s_\beta v^2}\{c_\beta[c_1 c_2
s_2M_{H_1^0}^2-c_2 s_3(c_1 s_2 s_3+s_1 c_3) M_{H_2^0}^2 \nn &&
+c_2
c_3(s_1 s_3-c_1 s_2 c_3) M_{H_3^0}^2] +s_\beta[s_1 c_2 s_2 M_{H_1^0}^2+c_2
s_3(c_1c_3-s_1s_2s_3)M_{H_2^0}^2
\nn
&& -c_2c_3(c_1s_3+s_1s_2c_3)M_{H_3^0}^2]\}\,,\label{lambdas}
\end{eqnarray}
with $s_\beta=\sin \beta$, $c_\beta=\cos \beta$ and $\mu$ is given by eq. (\ref{mu}).

The expressions  (\ref{lambdas}) are implemented in our numerical code.
Note that there are limits of non-CPV in
the C2HDM \cite{Per}, corresponding to the following values of the
$\a$-parameters:
\begin{eqnarray}
\a_2 &=& \pm \pi/2\,,\nonumber \\
\a_3 &=& \pm \pi/2\,, \nonumber \\
\a_2 &=&\a_3=0\,.
\end{eqnarray}
At these limits, the elements $R_{13}$, $R_{23}$ and $R_{33}$ of
the mixing matrix  (\ref{matRo}) vanish and the expression
 (\ref{M3}) becomes unstable. As this case is a subject of
a different model convention, namely it is already the CP conserving 2HDM,
our numerical code would produce an error message and interrupt the evaluation.

\section{Interaction Lagrangian}
\label{App:Lag}
For our calculations we have used the following part of the
C2HDM Lagrangian:

\subsubsection{Interactions of two quarks with a gluon}
\vspace*{-4mm}
\be
{\cal L}_{\bar{q} q g}&=&-g_s T^\alpha_{k l } G^\alpha_\mu
\bar{q}_k \gamma^\mu q_l\,, \quad k,l = 1,2,3\,, \quad \alpha =
1,...,8, \quad q = t,b\,,
\ee
where $T^\alpha/2$ are the Gell-Mann matrices and $g_s$ is the SU(3) strong coupling constant.

\subsubsection{Yukawa interactions of the neutral and the charged Higgses}
\vspace*{-4mm}
\begin{eqnarray}
 {\cal L}_{\bar{t} t H^0_j}&=&
\bar{t}(h^L_{t,j}P_L + h^R_{t,j}P_R)t H^0_j, \quad j =
1,2,3,\nonumber \\ h^L_{t,j}&=&-{1\over \sqrt{2}}(R_{j2}+i
c_\beta R_{j3}) h_t, \nonumber \\
h^R_{t,j}&=&-{1\over \sqrt{2}}(R_{j2}-i c_\beta R_{j3}) h_t,
\quad h_t~=~{g m_t\over \sqrt{2} m_W s_\beta}\, , \nonumber \\
{\cal L}_{\bar{b} b H^0_j}&=& \bar{b}(h^L_{b,j}P_L +
h^R_{b,j}P_R)b H^0_j, \quad j = 1,2,3,\nonumber \\
h^L_{b,j}&=&-{1\over \sqrt{2}}(R_{j1}+i s_\beta R_{j3}) h_b,
\nonumber \\ h^R_{b,j}&=&-{1\over \sqrt{2}}(R_{j1}-i s_\beta
R_{j3}) h_b, \quad h_b~=~{g m_b\over \sqrt{2} m_W c_\beta}\, ,\nonumber \\
{\cal L}_{t b H^\pm}&=&\bar{t}(y_t P_L +y_b P_R) b H^+
+\bar{b}(y_b P_L +y_t P_R) t H^-, \nonumber \\ y_t&=&h_t c_\beta,
\quad y_b~=~h_b s_\beta\, ,
\end{eqnarray}
with $s_\beta = \sin\beta$ and $c_\beta = \cos\beta$;
$P_L=(1-\gamma_5)/2$ and $P_R=(1+\gamma_5)/2$ denote the left and
right projection operators, respectively, $R_{jk}$ with $j,k=1,2,3$
are the elements of the mixing matrix ${\cal R}$, see
eq. (\ref{matRo}).

\subsubsection{Two quarks-W and two quarks-ghost interactions}
\vspace*{-4mm}
\begin{eqnarray}
{\cal L}_{t b W^\pm}&=&- {g\over \sqrt{2}}(\bar{t} \gamma^\mu P_L
b W^+_\mu + \bar{b} \gamma^\mu P_L t W^-_\mu ); \nonumber \\ {\cal
L}_{t b G^\pm}&=&\bar{t}(\tilde{y}_t P_L +\tilde{y}_b P_R) b G^+
+\bar{b}(\tilde{y}_b P_L +\tilde{y}_t P_R) t G^-, \nonumber \\
\tilde{y}_t&=&{g m_t\over \sqrt{2} m_W}, \quad \tilde{y}_b~=~{g
m_b\over \sqrt{2} m_W}\, .
\end{eqnarray}

\subsubsection{Triple scalar interactions with neutral and charged Higgses}
\vspace*{-4mm}
\begin{eqnarray}
 {\cal L}_{H^0_j H^+
H^-}&=&-{2 m_W \over g}(f_{H^0 H^+ H^-})_j
H^0_j H^+ H^- , \quad j=1,2,3,\nonumber \\ (f_{H^0 H^+ H^-})_j&=&
c_\beta[s^2_\beta (\l_1-\l_4-{\rm Re}(\l_5))+c^2_\beta \l_3] R_{j 1}+
\nonumber \\
&&s_\beta[c^2_\beta (\l_2-\l_4-{\rm Re}(\l_5))+s^2_\beta \l_3] R_{j 2}+
s_\beta c_\beta{\rm Im} (\l_5) R_{j 3}\, .
\end{eqnarray}

\subsubsection{Neutral Higgs-charged Higgs-W and
neutral Higgs-charged Higgs-ghost interactions}
\vspace*{-4mm}
\begin{eqnarray}
 {\cal L}_{H^0_j H^+ W^-}&=&{i g\over 2}(s_\beta R_{j 1}-c_\beta R_{j 2}+i R_{j
3})[H^0_j\delr H^+ W^{\mu-} - H^0_j \delr H^- W^{\mu+}
],\nonumber\\ &&j=1,2,3\, ,\nonumber \\
 {\cal L}_{H^0_j H^+ G^-}&=&{m_W\over
g}(f_{H^0 H^+ G^-})_j H^0_j
H^+ G^- + {\rm h.c.}, \quad j=1,2,3\,,\nonumber \\
(f_{H^0 H^+ G^-})_j&=&s_\beta [s^2_\beta (\l_4+{\rm Re}(\l_5))+c^2_\beta
(2 \l_1-\l_3-\l_{345})-i {\rm Im}(\l_5)] R_{j 1}+\nonumber \\
&& c_\beta [-c^2_\beta (\l_4+{\rm Re}(\l_5)) - s^2_\beta (2 \l_2-\l_3-
\l_{345})-i {\rm Im}(\l_5)] R_{j 2}+\nonumber \\
&& [i (\l_4- {\rm Re}(\l_5))+ (c^2_\beta -s^2_\beta) {\rm Im}(\l_5)] R_{j 3},
\nonumber \\
 \l_{345} &=& \l_3 +\l_4 + {\rm Re}(\l_5), \quad j=1,2,3\, .
\end{eqnarray}

\subsubsection{Interactions of a neutral Higgs with W-W, W-ghost
and ghost-ghost}
\vspace*{-4mm}
\begin{eqnarray}
{\cal L}_{H^0_j W^+W^-}&=&g m_W (c_\beta R_{j 1}+s_\beta R_{j 2})
H^0_j W^+_\mu W^{\mu -}, \quad j=1,2,3\, ,\nonumber \\ {\cal
L}_{H^0_j G^+W^-}&=&- {ig\over 2}(c_\beta  R_{j 1}+s_\beta R_{j
2}) [H^0_j\delr G^+ W^{\mu -}- H^0_j\delr G^- W^{\mu +} ], \quad
j=1,2,3\, ,
\nonumber \\ {\cal L}_{H^0_j G^+G^-}&=&-{2 m_W \over g
}[c_\beta (c^2_\beta \l_1+s^2_\beta \l_{345}) R_{j 1}+\nonumber \\
&& s_\beta (s^2_\beta \l_2+ c^2_\beta \l_{345}) R_{j 2}- s_\beta
c_\beta {\rm Im}(\l_5 )R_{j 3}] H^0_j G^+ G^-, \quad j=1,2,3\, .
\end{eqnarray}

\section{Implementation of the complex 2HDM into
FeynArts and FormCalc}\label{App:FAFC}

At present, there are several software packages for deriving Feynman
rules and doing calculations of particle processes on the market
 e.g. \cite{FeynArts,Oscar,feynrules}. For the purpose of our work it is convenient to
use the FA package. Up to now the FA package does not include
a model file for the 2HDM with CPV. In the
following sections we describe how we have implemented the new model file for calculations
in the C2HDM
into FA, and how we have extended the corresponding FC fortran drivers.

\subsection{FeynArts model file for the C2HDM}\label{App:feynarts}

At the moment, the diagram generator FA recognizes three
generic particle physics models: the SM, the MSSM and the 2HDM \cite{FeynArts}. The information about
the physics properties of each model (fields, their propagators
and their couplings) is collected in the corresponding model file.
The model files for these three models exist in two
different varieties: based only on the electroweak subset or
including in addition quantum chromodynamics: {\tt SM(QCD).mod},
{\tt MSSM(QCD).mod}, {\tt THDM(QCD).mod}. All couplings are expressed in
terms of the parameters of the relevant Lagrangian.

The 2HDM model implemented in FA is based on the Higgs
potential  (\ref{CTHDMpot}), but with $m_{12}^2=\l_{6}=\l_7=0$ and all other
parameters are real. The latter constraints refer to a model with
the particle content of the SM, but a physical Higgs sector
analogous to the one of the MSSM, see e.g. \cite{CPcons}.
Usually, this is the most commonly used version studied in literature,
also called the CP-conserving 2HDM due to absence of complex
parameters. (Note that there exist another public software package
for calculations in the CP-conserving 2HDM \cite{Oscar}.)
In order to generalize the existing 2HDM model file for the general complex
case, which is described by the Higgs potential  (\ref{CTHDMpot}),
one should recalculate the couplings in terms of the full set
of parameters in eq. (\ref{CTHDMpot}).\footnote{In this case the fields
and their propagators are not affected.} The new FA model
file {\tt CTHDM.mod} consists of 270 couplings, which includes all Higgs interactions:
\begin{itemize}
\item Higgs-vector boson interactions extracted from the covariant
derivatives  (\ref{covder})
\item  Triple and quartic Higgs self-interactions from the scalar potential
 (\ref{CTHDMpot})
\item  Higgs interactions  with Fadeev-Popov ghosts
\item  Yukawa interactions\footnote{Note that
here we assume one Higgs doublet to couple with only up-type
fermions and the other one- only with down-type fermions or the so
called 2HDM type II.}
\end{itemize}
Some of them are obtained explicitly e.g. in \cite{Per,
Hcoupl}. The rest of the couplings needed to complete the
model file are not concerned and we have taken them from the already existing
model file {\tt THDM.mod}. These are: vector boson self-interactions,
fermion-vector interactions, and with QCD couplings: gluon
self-interactions, gluon-ghost and gluon-quark interactions.
The model file can be used independently
for diagram generation in the general complex case of the 2HDM and
it is not necessarily related to further calculations. There is a
rule included for a switch to the case $\l_{6}=\l_7=0$, which describes the 2HDM with a softly
broken $Z_2$ symmetry  (\ref{THDMpot}).
In order to go back to the CP-conserving case one must also set $m_{12}^2$ to zero and consider
$\l_5$ as a real parameter. The complete {\tt CTHDM.mod} is
too lengthy to be printed explicitly in this note. It can be
found and downloaded from the FA website:
{$www.feynarts.de$}.

\subsection{FormCalc drivers for the C2HDM}
\label{App:formcalc}

After the diagram generation with the new FA model file  {\tt CTHDM.mod} FC calculates
the squared matrix elements with the help of
Form \cite{Form} and the resulting expressions are translated into Fortran for
the further numerical evaluation.
For consistency, the Fortran drivers necessary for the initialization of the
model have to be extended to include the new set of parameters of
the Higgs potential, the relations between them and the
constraints on them.
We would like to note that in spite
of the fact that the new FA model file is written
for the most general complex case with scalar potential  (\ref{CTHDMpot}),
the extension of the FC fortran drivers
is made only for the case of a softly broken
$Z_2$ symmetry of the 2HDM Lagrangian, described by the scalar potential  (\ref{THDMpot}).
The existing initialization file {\tt model\_thdm.F}
is replaced by the new file  {\tt model\_cthdm.F}, which defines all
parameters of the potential  (\ref{THDMpot}),
the theoretical and experimental constraints on them. Also the set
of input parameters for the numerical evaluation in the
C2HDM is defined therein. All implemented expressions are listed and discussed in the present
paper.

Since the triple and the quartic scalar couplings have
complicated and lengthly expressions, during the calculation
of a given process they are replaced by
generic couplings named
{\tt cS(i,j,k)} and {\tt qS(i,j,k,l)}.
Furthermore, all triple and quartic scalar couplings are evaluated in
the  {\tt model\_cthdm.F} file.

\subsubsection{Inputs}
Based on the case with a softly broken $Z_2$ symmetry of the 2HDM Lagrangian described by the
scalar potential  (\ref{THDMpot})
we work with the following set of input parameters:
\begin{itemize}
\item ${\tt Mh0}=M_{H_1^0}$ -- mass of $H_1^0$
\item ${\tt MHH}=M_{H_2^0}$ -- mass of $H_2^0$
\item ${\tt MHp}=M_{H^+}$ -- mass of the charged Higgs
\item ${\tt TB}=v_1/v_2$ -- the ratio of the VEVs
\item ${\tt rm12}={\rm Re}\,(m_{12})$ -- real part of the $m_{12}$ parameter of the Higgs potential
\item ${\tt alp1}=\alpha_1$ -- mixing angle
\item ${\tt alp2}=\alpha_2$ -- mixing angle
\item ${\tt alp3}=\alpha_3$ -- mixing angle
\item ${\tt rm0}=M_0=120$ GeV: reference point to substract the SM Higgs contribution from $\Delta \rho$, see section \ref{App:rho}
\end{itemize}
Once the input parameters are initiated in {\tt run.F}, in {\tt model\_cthdm.F} the mass of the heaviest neutral Higgs $M_{H_3^0}$
is calculated by using eq. (\ref{M3}). Then $M_{H_3^0}$ has to satisfy the following conditions:\\
i) $M_{H_3^0}^2>0$, which means that $H_3$ is not a tachyonic mode;\\
ii) $M_{H_1^0} \leq M_{H_2^0} \leq M_{H_3^0}$, see section \ref{sec:genpot}.\\
When an allowed value for $M_{H_3^0}$ is obtained, the code proceeds with the
evaluation of the $\lambda_i$ as given in appendix~\ref{App:lambda}.
Finally, the triple and the quartic scalar couplings are calculated in {\tt model\_cthdm.F} .

\subsubsection{Constraints}
The file {\tt model\_cthdm.F} contains the following constraints, see section \ref{sec:th.constr.} and section \ref{sec:exp.constr.}:
\begin{itemize}
\item Vacuum stability
\item Perturbativity and unitarity
\item $\Delta\rho$ constraint
\end{itemize}
All constraints can be switched on and off. In addition, the upper and the lower $\Delta\rho$ bounds
can be modified easily.


\acknowledgments
The authors thank Dietrich Liko for his help in Grid computing and
Walter Majerotto for useful comments. A. Arhrib and E. Christova thank E. Ginina and H. Eberl for the
hospitality during their visits in Vienna. The authors acknowledge support from EU under the MRTN-CT-2006-035505 network
programme. This work is supported by the "Fonds zur F\"orderung der
wissenschaftlichen Forschung"  (FWF) of Austria, projects No. P18959-N16 and I297-N16. The work of E. Christova and
E. Ginina is supported by the Bulgarian National Science Foundation, grant 288/2008.
%


%

\end{document}